\documentclass{article}
\usepackage{amssymb}
\usepackage{cite,multirow} 
\usepackage[dvips]{graphicx,epsfig}
\usepackage{graphics}
\def\1ad{\mbox{\normalsize $^1$}}
\def\2ad{\mbox{\normalsize $^2$}}
\def\3ad{\mbox{\normalsize $^3$}}
\def\4ad{\mbox{\normalsize $^4$}}
\def\5ad{\mbox{\normalsize $^5$}}
\def\6ad{\mbox{\normalsize $^6$}}
\def\7ad{\mbox{\normalsize $^7$}}
\def\8ad{\mbox{\normalsize $^8$}}

\makeatletter
\@addtoreset{equation}{section}
\makeatother
\renewcommand{\theequation}{\thesection.\arabic{equation}}
 
\def\beq{\begin{equation}}                     %
\def\eeq{\end{equation}}                       %
\def\bea{\begin{eqnarray}}                     
\def\eea{\end{eqnarray}}                       
\def\0 {\nonumber}

\begin{document}

\setcounter{page}{0}
\begin{titlepage}
\titlepage
\rightline{hep-th/0505220}
\rightline{Bicocca-FT-05-11}
\vskip 3cm
\centerline{{ \bf \Large The dual superconformal theory for $L^{p,q,r}$ 
manifolds}}
\vskip 1cm
\centerline{{\bf Agostino Butti, Davide Forcella
 and Alberto Zaffaroni}}
\vskip 1.5truecm
\begin{center}
\em 
$^a$ Dipartimento di Fisica, Universit\`{a} di Milano-Bicocca \\ 
P.zza della Scienza, 3; I-20126 Milano, Italy\\
\vskip .4cm

\vskip 2.5cm
\end{center}
\begin{abstract}
We present the superconformal gauge theory living on the world-volume
of D3 branes probing the toric singularities with horizon the 
recently discovered Sasaki-Einstein manifolds $L^{p,q,r}$.
Various checks of the identification are made by comparing the
central charge and the R-charges of the chiral fields with the information that
can be extracted from toric geometry. Fractional branes
are also introduced and the physics of the associated duality cascade
discussed.

\vskip1cm

\end{abstract}
\vskip 0.5\baselineskip

\vfill
 \hrule width 5.cm
\vskip 2.mm
{\small
\noindent agostino.butti@mib.infn.it\\
davide.forcella@mib.infn.it\\
alberto.zaffaroni@mib.infn.it}
\begin{flushleft}
\end{flushleft}
\end{titlepage}
\large
\section{Introduction}
D3 branes living at conical Calabi-Yau 
singularities are a good laboratory for the AdS/CFT correspondence since 
its early days. The world-volume
theory on the branes is dual to a type  IIB background
of the form $AdS_5\times H$, where $H$ is the horizon manifold 
\cite{kw,horizon}. The correspondence between conical singularities
and superconformal gauge theories  has also given information 
on non-conformal ones. One of the few known examples 
of regular backgrounds dual to confining gauge theories,
the Klebanov-Strassler solution \cite{ks}, is indeed
obtained by introducing fractional branes at a conifold singularity.
In the conformal case, supersymmetry requires that $H$ is a Sasaki-Einstein 
metric. Until few months ago, the only known Sasaki-Einstein metrics
were the round sphere $S^5$ and $T^{1,1}$, the horizon of the conifold.
Recently, an infinite class of new regular Sasaki-Einstein metrics 
with $SU(2)\times U(1)\times U(1)$ isometry was
constructed \cite{gauntlett}. These manifolds are labeled by
two integers $\bar p$ and $\bar q$ and have been named $Y^{\bar p,\bar q}$.
With the determination of the corresponding dual gauge theory \cite{benvenuti},
new checks of the AdS/CFT correspondence were possible \cite{bertolini,benvenuti,kleb,kru}. As well known,
the central charge of the CFT and the dimension of some operators can be
compared with the volumes of $H$ and of some of its submanifolds.
In particular, the a-maximization technique \cite{intriligator} now
allows for a detailed computation of the relevant quantum field theory
quantities. Needless to say, the agreement of the two computations
is perfect. 

More recently, a generalization of the $Y^{\bar p,\bar q}$ 
manifolds with smaller isometry $U(1)^3$ and depending on 
three integers $p,q,r$ has been constructed \cite{CLPP,MSL}. 
When $p,q$ and $r$ satisfy some conditions, these Sasaki-Einstein manifolds, 
named $L^{p,q,r}$,
are smooth. 
In this paper, we will construct the dual superconformal
gauge theory, using a powerful method developed in \cite{dimers}. 
The precise correspondence between conical Calabi-Yau singularities
and superconformal gauge theories is still unknown. However,
a remarkable progress has been recently made for the class of
Gorenstein toric singularities. The brane tiling (dimers) construction
\cite{dimers}, an ingenious generalization of the Brane Boxes \cite{boxes,boxes2},
introduces a direct relation between an Hanany-Witten realization \cite{hw}
for gauge theory and the toric diagram. In particular, from the quiver
associated with a superconformal gauge theory one
can determine the dual brane tiling configuration, a dimer lattice.
It is then possible to associate a toric diagram with each of these lattices, 
identifying the dual Calabi-Yau. The inverse process
(to associate a gauge theory with a given singularity) is more difficult. 
Hopefully, in the long period, the dimers technology will allow to
define a one-to-one correspondence between CFTs and toric singularities.
For the moment, we will determine the superconformal gauge theory
by using some analogy with the $Y^{\bar p,\bar q}$ case. 

To implement the dimer technology, we need the toric description of
the cone over the $L^{p,q,r}$ manifolds. The construction of the
toric diagram for all integers $p,q,r$ and the toric description
of the Calabi-Yau is described in the first part of this paper
using results given in \cite{CLPP,MSL}. We will also present
explicit formulae for the volume of $L^{p,q,r}$ and 
the volumes of special submanifolds. With the knowledge of the toric diagram,
we can engineer a dimer configuration that is general enough to
contain all the infinite theories with diagram associated with the 
numbers $p,q,r$. We are helped by the fact that the toric diagram has only
four external vertices. In the second part of the paper, we discuss
the dual gauge theory. We compute the central charge and the
dimensions of dibaryons using the a-maximization technique \cite{intriligator}
and we compare these results with the AdS/CFT predictions based on volumes.
We find a complete agreement. 

Even if expected, the agreement always contains a little bit of magic.
In particular, we have two different algebraic procedures for computing
the R-symmetry charges of the fields and the volumes. The first is based
on the maximization of the central charge. The second one can be
efficiently encoded in a geometrical minimization procedure for determining
the Reeb vector discovered in \cite{MSY}. The two procedures deal with
different test quantities (the R-charges on one side and the components
of the Reeb vector on the other) and with different functions to be extremized.
It is remarkable that the two methods give exactly equivalent results.
In the case of $L^{p,q,r}$ we are not able to analytically solve the 
extremization problems. Fortunately, the two procedures can be easily 
automatized in Mathematica. The
numerical coincidence of the results for all $p,q,r$ is indeed impressive.
   
Notice that the agreement of results in the gauge theory and the supergravity
side can be regarded not only 
as another non-trivial check of the AdS/CFT 
correspondence, but also as a check of the brane tiling construction 
\cite{dimers}. 
The latter is indeed still at the level of a conjecture. We are quite confident
that the conjecture will become soon an established result and will lead
to an important improvement of our knowledge in the field.

We also discuss the inclusion of fractional branes. As usual this leads
to a duality cascade. Unfortunately, as in the $Y^{\bar p,\bar q}$ case, it is
most plausible that the theories associated with smooth $L^{p,q,r}$
have no
supersymmetric vacuum. We briefly discuss the arguments leading 
to these conclusions \cite{unpublished,berenstein,hananyuranga,mavasesipuo!}.

The paper is organized as follows. In Section \ref{geometry} we review the formulae
for the metric of $L^{p,q,r}$. Section \ref{toric} describes the toric geometry
of the cones over $L^{p,q,r}$ determining the toric diagram. We also 
discuss the regularity of the metric for all values of $p,q,r$ using
toric geometry. We set up a minimization problem for
the determination of the Reeb vector following the results in \cite{MSY}
and we give closed expressions for the volume of the submanifolds associated
with the edges of the toric fan. In Section \ref{pregauge} we discuss
the tiling. In Section \ref{gauge} we describe the dual
gauge theory and perform the a-maximization. 
In Section \ref{fractional} 
we discuss the inclusion of fractional branes in the game
and the related duality cascade. Finally, in the Appendix we give details
on the brane construction and provide specific examples.
    
While finishing this work, a paper with partial overlap appeared \cite{kru2}.
It also presents the central charge and the R-charges of fields.
Similar results have been obtained also in \cite{tomorrow}.

\section{The $L^{p,q,r}$ spaces }\label{geometry}

Recently a new infinite class of five-dimensional Sasaki-Einstein
spaces was discovered \cite{CLPP,MSL}. They were denoted $L^{p,q,r}$.
In this Section, we briefly review the geometry of these spaces.
Elaborating on \cite{CLPP,MSL}, we give explicit formulae for 
the metric of these spaces and collect expressions that can be useful
for the evaluation of volumes. 
 
Our starting point is a set of D3-branes probing a CY conical singularity
\cite{kw}. The geometry of the transverse space is a cone with $L^{p,q,r}$
base
\begin{equation}
ds^2_{{\rm CY}}=dr^2+r^2 ds^2_{(L^{p,q,r})}
\end{equation}

The smooth $L^{p,q,r}$ spaces are characterized by 
three coprime  positive integers $p$, $q$, $r$ with $0<p\leq q$, $0< r < p+q $.
The metrics have $U(1) \times U(1) \times U(1)$ isometry enlarging to $SU(2) \times U(1) \times U(1)$ in the special case $p+q = 2r$, which reduces to the previously-known spaces  $ Y ^{\bar{p},\bar{q}}= L^{\bar{p}-\bar{q},\bar{p}+\bar{q},\bar{p}}$ 
\cite{gauntlett}. The topology of these spaces is $S^3\times S^2$.

For Sasaki-Einstein spaces one can write the local form of the metric as 
follows: 
\begin{equation}\label{metrica}
ds_{(L^{p,q,r})}^2 = (d\tau + \sigma )^2 +ds_4 ^2
\end{equation}
In the case of the spaces $L^{p,q,r}$
\begin{eqnarray}
\label{quattro}
ds_4 ^2 &=&  \frac{\rho ^2 dx^2}{4\Delta _x} + \frac{\rho ^2 d\theta ^2}{\Delta _{\theta}} + \frac{\Delta _x }{\rho ^2} \big( \frac{ \sin ^2 \theta }{\alpha}d\phi + \frac{\cos ^2 \theta }{\beta}d\psi \big)^2\nonumber  \\ 
&+& \frac{\Delta _{\theta} \sin ^2 \theta \cos ^2 \theta }{\rho ^2}\big(\frac{\alpha - x }{\alpha}d\phi - \frac{\beta - x }{\beta}d\psi \big)^2
\end{eqnarray}
where
\begin{eqnarray}\label{sigma}
\sigma &=& \frac{(\alpha - x)\sin ^2 \theta }{\alpha}d\phi + \frac{(\beta - x )\cos ^2 \theta }{\beta}d\psi \nonumber\\
\Delta _x &=& x(\alpha - x)(\beta - x ) - \mu \nonumber\\
\Delta_{\theta} &=& \alpha \cos ^2 \theta + \beta \sin ^2 \theta\nonumber\\
\rho ^2 &=& \Delta _{\theta} - x
\end{eqnarray}

We can set one of the parameters $\alpha$, $\beta$, $\mu$ to any non-zero constant  by rescaling the other two and $x$. Hence the metrics depend on two non-trivial parameters, which we will take to be $\beta$ and $\mu$. 

The metrics are in general of cohomogeneity 2, with toric principal orbits $U(1)\times U(1) \times U(1) $ . 
In order to obtain metrics on complete non-singular manifolds we have to take the following range of variables: $0< \theta < \pi /2 $, $x_1 < x < x_2$, where $x_1 $ and $x_2$ are the two lowest roots of $\Delta _x =0$. The ranges of the coordinates $\phi$, $\psi$, $\tau$ as well as the toric $T^3$ action are described in the next Section. In this range there are four degeneration surfaces respectively at $\theta =0$, $\theta = \pi /2 $, $ x=x_1 $, $ x=x_2 $ where, in order, the normalized four Killing vectors 
vanish:  
\begin{equation}
\label{uno}
\frac{\partial}{\partial \phi},\qquad 
\frac{\partial}{\partial \psi},\qquad 
l_i = a_i\frac{\partial}{\partial \phi}+ b_i\frac{\partial}{\partial \psi}+c_i\frac{\partial}{\partial \tau} 
\end{equation}
where $i=1,2$ and
\begin{eqnarray}\label{ai}
a_i &=& \frac{\alpha c_i}{x_i - \alpha}\nonumber\\ 
b_i &=& \frac{\beta c_i}{x_i - \beta} \nonumber\\
c_i &=& \frac{(\alpha - x_i)(\beta - x_i)}{2(\alpha + \beta )x_i - \alpha \beta - 3 x_i ^2} 
\end{eqnarray}
Notice that these quantities satisfy the equation 
\begin{equation}\label{aibici}
1 + a_i + b_i + 3c_i = 0 
\end{equation}
To obtain a non singular manifold, we have to impose a linear relation among the Killing vectors  
\begin{equation}\label{reluno}
p\,l_1 + q\,l_2 + r\frac{\partial}{\partial \phi} + s\frac{\partial}{\partial \psi} = 0 
\end{equation}
with $(p,q,r,s)$ four coprime integers with  
\begin{equation}\label{reldue}
p + q = r + s 
\end{equation}
Notice that all triples  chosen from $(p,q,r,s)$ also are coprime.
In order to have smooth geometry in 5-d the parameters need to obey $\alpha , \beta \geq x_2 $ with $x_3 \geq x_2 \geq x_1 \geq 0 $, which imply $ q \geq p > 0 $
 and $p+q > r > 0$.

If we set $ p+q=2r $ $\Rightarrow$ $r=s$ $\Rightarrow$ $\alpha = \beta $ it easy to see that the $L^{p,q,r}$ spaces reduce to the $Y^{\bar p,\bar q}$ 
spaces with the relations:
\begin{equation}\label{p}
\bar{p} - \bar {q} = p, \qquad \bar{p} + \bar {q} = q,\qquad
\bar{p} = r 
\end{equation}
For reader convenience we exhibit the transformation laws that convert
the metric to the one commonly used for the $Y^{\bar p,\bar q}$ spaces:
\begin{equation}\label{psi}
\bar{\psi}' = 3\tau + \psi + \phi,\,\,\, 
\bar{\phi} = \phi - \psi ,\, \,\,
\bar{\beta} = - \phi - \psi, \,\, \,
\bar{\theta} = 2\theta, \, \,\,
\bar{y} = \frac{3x - \alpha}{2\alpha}
\end{equation}
The $L^{p,q,r}$ metric indeed reduces to 
\begin{eqnarray}\label{ypq}
ds_{(Y^{\bar{p},\bar{q}})}^2 &=& \frac{1-\bar{y}}{6}(d\bar{\theta} ^2 + \sin ^2 \bar{\theta} d\bar{\phi}^2) + \frac{d\bar{y}^2}{w(\bar{y})q(\bar{y})}+ \frac{1}{36}w(\bar{y})q(\bar{y})(d\bar{\beta} + \cos \bar{\theta} d\bar{\phi})^2 \nonumber\\ &+& \frac{d\bar{\psi}'}{3} + \frac{1}{3}(-\cos \bar{\theta}d\bar{\phi} + \bar{y}(d\bar{\beta} + \cos \bar{\theta}d\bar{\phi}))]^2
\end{eqnarray}
that coincides with the $Y^{\bar{p},\bar{q}}$ metric with 
\begin{eqnarray}\label{wy}
w(\bar{y}) &=& \frac{2(a- \bar{y} ^2)}{1 - \bar{y}}\nonumber\\
q(\bar{y}) &=& \frac{a- 3 \bar{y} ^2 +2\bar{y}^3}{a - \bar{y}^2}
\end{eqnarray}
and $a$ related to $\mu$ by
\begin{equation}
\mu = \frac{4}{27}(1-a)\alpha^3
\end{equation}

We can find expressions for $x_1$, $x_2$, $\alpha $, $\beta$, in terms of $p$, $q$, $r$, $s$.
Using the equations (\ref{aibici}, \ref{reldue}) and 
\begin{eqnarray}\label{pcuno}
pc_1 +qc_2 &=& 0 \nonumber\\
pa_1 + qa_2 + r &=& 0
\end{eqnarray}
we obtain \footnote{These equations are strictly valid in the case $r \neq s$.}
\begin{eqnarray}\label{x1}
x_1 &=& \frac{-rs\alpha ^2 + p(s-r)\alpha \beta + sr\beta ^2 + (\alpha -\beta)\sqrt{\delta}}{(p-3r)s\alpha + (3s-p)r\beta}   \nonumber\\
x_2 &=& \frac{-rs\alpha ^2 + q(s-r)\alpha \beta + sr\beta ^2 + (\beta -\alpha)\sqrt{\delta}}{(q-3r)s\alpha + (3s-q)r\beta}   
\end{eqnarray}
with
$$\delta=sr(r(q-r)(\alpha ^2 - \alpha \beta + \beta ^2) + p(r\alpha ^2 - (q+r)\alpha \beta +r\beta ^2))$$
To find $\alpha $, $\beta$ we use the relations
$\Delta  _{x_1} =  \Delta _{x_2} = 0$.
By eliminating $\mu$ from the first relation we obtain
\begin{equation}\label{x1x2}
x_1(\alpha - x_1)(\beta - x_1 ) = x_2(\alpha - x_2)(\beta - x_2 )  
\end{equation}  
and using now equation~(\ref{x1}) we find the quartic equation
\begin{eqnarray}\label{quartic}
&&s^3 r \alpha ^4 - s^2((q-3r)r + p(p+r))\alpha ^3 \beta + 2sr(pq - 2(p+q)r +2r^2 ) \alpha ^2 \beta ^2\nonumber\\
 &&+ r^2(2p^2 + 3pq + 2q^2 - 5(p+q)r + 3r^2)\alpha \beta ^3 + sr^3 \beta ^4 =0 
\end{eqnarray}  
This can be solved to give a long expression for $\beta / \alpha $. 
If we use $\beta$, $\mu$ as free parameters in the metric (\ref{metrica}) we can set $\alpha = 1$ and so we have the expressions for $x_1$, $x_2$, $\alpha $, $\beta$, $\mu$ in function of $p$, $q$, $r$, $s$.
We must verify that the solution of equation~(\ref{quartic}) gives rise to
positive definite metrics. This implies $q \geq p$ \cite{CLPP}.
Using the metric (\ref{metrica}) it is easy to find the volume of the $L^{p,q,r}$ spaces 
\begin{equation}\label{volume}
V = \frac{\pi ^3 |c_1|(x_2 - x_1)(\alpha + \beta - x_1 - x_2 )}{\alpha \beta q}
\end{equation}  
Each variable in (\ref{volume}) is known as a function of $p$, $q$, $r$.
Therefore, if we like, we can write a (very long) closed equation for the volume in function of $p$, $q$, $r$.   

\section{The toric diagram for $L^{p,q,r}$}
\label{toric}
In this Section we build the toric diagram for the CY singularities
with horizon $L^{p,q,r}$. The impatient reader can find all the relevant
information about the toric diagram in Figure 1. 
We also compute the 
volumes of special submanifolds of $L^{p,q,r}$, using a method 
developed in \cite{MSY}. These results will be later compared with those
of a-maximization. We refer to \cite{MS} for a review of all the
necessary notions of toric geometry used in this Section. 

\subsection{Building the toric diagram}
In the previous Section we reviewed the local form of the metric of the Calabi-Yau cones over the Sasaki-Einstein spaces $L^{p,q,r}$; the toric action $U(1)^3$ on such cones is locally given by the vectors fields: $\frac{\partial}{\partial \phi}$, $\frac{\partial}{\partial \psi}$, $\frac{\partial}{\partial \tau}$.

As reviewed in \cite{MS}, on (six dimensional) symplectic toric cones $M$ we can define a moment map $\mu$: $M \rightarrow R^3$ whose image is a three dimensional strictly convex rational polyhedral cone ${\cal C}$ 
(the dual fan, in the language of toric geometry \cite{fulton}). Moreover the original CY cone is reconstructed as a $T^3$ fibration over the polyhedral cone, with one or two cycles of the $T^3$ degenerating over faces or edges respectively of the dual fan.
In our case, $L^{p,q,r}$, we know that the image of the momentum map is a polyhedral cone with four faces, since there are only four subvarieties with a degenerating Killing vector field \cite{CLPP}. Such divisors are the submanifolds at $\theta=0$, $\theta=\pi/2$, $x=x_1$, $x=x_2$, with degenerating vectors given respectively by $\frac{\partial}{\partial \phi}$, $\frac{\partial}{\partial \psi}$, $l_1$, $l_2$ in formula~(\ref{uno}). With the normalizations of the previous Section, the orbits of these vector fields must be closed with period $2\pi$ in order to avoid conical singularities when one approaches the faces of the polyhedral cone \cite{CLPP}. 

The $T^3$ fibered over the polyhedral cone is described locally by the coordinates ($\phi$, $\psi$, $\tau$), but we have not yet given a global description of the $T^3$ action and of the closed orbits. For example, we expect that in general the orbits of the Reeb vector $\frac{\partial}{\partial \tau}$ will not be closed except for the regular (or quasi-regular) case \cite{MS}. 
The toric action can be described giving three identifications in the space ($\phi$, $\psi$, $\tau$) along three independent directions, or equivalently 
giving a basis of Killing vectors $e_1$, $e_2$, $e_3$ that describe an effectively acting $T^3$ \cite{MS}. $e_1$, $e_2$, $e_3$ will be a linear combination of $\frac{\partial}{\partial \phi}$, $\frac{\partial}{\partial \psi}$, $\frac{\partial}{\partial \tau}$ and we understand that they are normalized such that their orbit is closed with period $2 \pi$. Of course any $SL(3,Z)$ combination of the $e_i$ will also be a good basis; as usual, the toric diagram is
only defined up to $SL(3,Z)$ transformations \cite{fulton}.
Notice that we are building the $T^3$ action so as to obtain a regular CY cone at $r\neq 0$. 

The degenerating vectors can be written as linear combinations of the $e_i$:
\begin{equation}
\left(
\begin{array}{c}
l_1 \\ 
l_2 \\
\frac{\partial}{\partial \phi}\\ 
\frac{\partial}{\partial \psi}
\end{array}
\right)
=
\left(
\begin{array}{ccc}
A_{11} & A_{12} & A_{13}\\
A_{21} & A_{22} & A_{23}\\
A_{31} & A_{32} & A_{33}\\
A_{41} & A_{42} & A_{43}\\
\end{array}
\right)
\left(
\begin{array}{c}
e_1 \\
e_2 \\
e_3 \\
\end{array}
\right)
\end{equation} 

Now notice that the four vectors orthogonal to the faces of the polyhedral cone are just the killing vectors that vanish on such faces \cite{MS}; the coordinates of such vectors in the basis $e_i$ are just the integers vectors in $R^3$ that describe the toric diagram (fan). Therefore in our notation the fan is simply generated by the four lines of the matrix $A_{ij}$. 
To avoid additional singularities the coefficients $A_{ij}$ must be integers and each triple $A_{ij}$ with fixed $i$ must be coprime,
so that the fan is generated by primitive vectors.
Up to now there is only a possible ambiguity in the choice of signs for the vectors of the fan; conventionally they are all inward pointing.
We now use the fact that for a CY cone we can perform an $SL(3,Z)$ transformation to set the first coordinate of the vectors of the toric diagram equal to one, that is
 $A_{1,j}=\pm 1$. But then the requirement~(\ref{reluno}) 
\begin{equation}
p l_1 +q l_2 +r \frac{\partial}{\partial \phi} + s\frac{\partial}{\partial \psi}=0
\label{linear}
\end{equation}
and the equality $p+q=r+s$ gives us the choice $A_{11}=A_{12}=-1$, and $A_{13}=A_{14}=1$. At this point we can perform another $SL(3,Z)$ transformation to set
 the remaining two components of, say, $l_1$ to $(0,0)$. Our matrix becomes: 
\begin{equation}
\left(
\begin{array}{c}
-l_1 \\ 
-l_2 \\
\frac{\partial}{\partial \phi}\\ 
\frac{\partial}{\partial \psi}
\end{array}
\right)
=
\left(
\begin{array}{ccc}
1 & 0 & 0\\
1 & P & Q\\
1 & R & S\\
1 & F & G\\
\end{array}
\right)
\left(
\begin{array}{c}
e_1 \\
e_2 \\
e_3 \\
\end{array}
\right)
\end{equation} 

In the physics literature, it is common to give, instead of the 
three-dimensional vectors defining the fan, the toric diagram consisting
in the projection of the vectors on the plane where they all live.
In this case, the toric diagram is given by ignoring the first coordinate and consists of the points: $(0,0)$, $(R,S)$, $(P,Q)$, $(F,G)$.

Let us study the smoothness of the CY cone.
Consider the face $x=x_1$ where $l_1$ degenerates; the $T^3$ fibration over this face reduces to a $T^2$ fibration and the corresponding lattice is obtained by the lattice $(e_1,e_2,e_3)$ taking the quotient along the direction $l_1$, which corresponds to $e_1$. Therefore over this face the $T^2$ lattice is simply the plane $(e_2,e_3)$, and the projected vectors $-l_2$, $\frac{\partial}{\partial \phi}$, $\frac{\partial}{\partial \psi}$ are $(P,Q)$, $(R,S)$, $(F,G)$. The face $x=x_1$ intersects also the two faces corresponding to $\theta=0$ and $\theta=\pi/2$, and to avoid conical singularities along the two edges where $x=x_1$ intersects these faces we must impose $hcf(R,S)=1$ and 
$hcf(F,G)=1$. Note that regularity does not require $hcf(P,Q)=1$ since obviously the divisor $x=x_1$ does not intersect $x=x_2$.

Since $R$ and $S$ are coprime, it is always possible to find an $SL(2,Z)$ transformation that sends (R,S) in (1,0). With this choice the two remaining components of equation (\ref{linear}) read:
\begin{eqnarray}
q \, Q = s \, G \label{eq1}\\
s \, F-q \, P+r=0
\label {eq2}
\end{eqnarray}

We have to impose the regularity conditions at the two remaining edges of the the dual fan; by taking the appropriate quotients along the degenerating directions in the lattice
 $(e_1,e_2,e_3)$ it is easy to derive the conditions $hcf(P-1,Q)=1$ and $hcf(F-P,G-Q)=1$, together with the old one $hcf(F,G)=1$. Indeed we have rederived the well known result in toric geometry that regularity is assured iff the four sides of the toric diagram $(0,0)$, $(1,0)$, $(P,Q)$, $(F,G)$ do not pass through integers points \cite{fulton}.

Let's choose $(p,q,r,s)$ coprime: we can always divide the defining equation (\ref{linear}) by a possible 
common factor; from $p+q=r+s$ it follows that every sub-triple of
$(p,q,r,s)$ is coprime. Then equation (\ref{eq2}) says that
$hcf(q,s)=1$ otherwise a common factor should also divide $r$, and therefore also $p=r+s-q$. But if $q$ and $s$ are coprime the general solution of (\ref{eq1}) is 
$Q=n \, s$ and $G=n \, q$ with $n$ an arbitrary integer $n \in \mathbb{Z}$.

Let us concentrate on the ``minimal'' regular case $n=1$: $Q=s$, $G=q$. The regularity condition $hcf(F,G)=1$ is satisfied iff $hcf(q,r)=1$: suppose that there is an integer 
$k$ that divides $q$ and $r$; then by (\ref{eq2}) it divides also the product $s F$, but since it cannot divide $s$, it must divide $F$. So $k$ would divide $F$ and $G=q$. The converse is also true. Analogously it is straightforward to show, using (\ref{eq2}), that the other regularity conditions are satisfied 
iff $hcf(p,r)=1$ and $hcf(p,s)=1$.

We have to determine the two remaining parameters $F$ and $P$ in order 
to complete the description of the toric diagram. Equation ($\ref{eq2}$) has infinite solutions. In fact it is a well known result in linear Diophantine equations that when $hcf(q,s)=1$ equation (\ref{eq2}) always admits solutions, with the most general of the form:
\begin{equation}
\begin{array}{c@{\hspace{5em}}c}
\left\{
\begin{array}{c}
F=F_0 +\lambda \, q\\
P=P_0 +\lambda \,s
\end{array}
\right.
&
\lambda \in \mathbb{Z}
\end{array}
\label{diophantine}
\end{equation}
where $F_0$ and $P_0$ are particular solutions.
There is not an explicit formula for $F_0$ and $P_0$ in terms of $q,r,s$, but of course there is an algorithm to determine the solution given the inputs $q,r,s$.
The important point is that all the solutions (\ref{diophantine}) describe the same toric geometry, since there is an $SL(2,\mathbb{Z})$ transformation that sends a given solution $(0,0)$, $(1,0)$, $(P_0,s)$, $(F_0,q)$ into any other of the family:
\begin{equation}
\left(
\begin{array}{cc}
1 & \lambda\\
0 & 1
\end{array}
\right)
\left(
\begin{array}{c}
P_0\\
s
\end{array}
\right)=
\left(
\begin{array}{c}
P_0+\lambda \,s\\
s
\end{array}
\right) \, , \hspace{2em}
\left(
\begin{array}{cc}
1 & \lambda\\
0 & 1
\end{array}
\right)
\left(
\begin{array}{c}
F_0\\
q
\end{array}
\right)=
\left(
\begin{array}{c}
F_0+\lambda \,q\\
q
\end{array}
\right)
\label{class}
\end{equation}

To summarize, all the smooth $L^{p,q,r}$ are characterized by four coprime positive integers $(p,q,r,s)$ such that $p+q=r+s$ and $hcf(p,r)=hcf(p,s)=hcf(q,r)=hcf(q,s)=1$; remember also from the previous Section
that a positive definite and well defined metric requires $q\geq p$. For this reason, in the following we will also denote these spaces with $L^{p,q;r,s}$.
Under these conditions we can define periodicities to get a smooth CY cone with the ``minimal'' choice $n=1$; the toric diagram is given by the points
\begin{equation}\label{to}
(0,0),\qquad (1,0),\qquad (P,s),\qquad (F,q)
\end{equation}
\begin{figure}[h!!]
\begin{minipage}[t]{\linewidth}
\begin{minipage}[t]{0.45\linewidth}
\vspace{0pt} \verb| | \\
\vspace{1.5em}
\[
\begin{array}{c@{\rightarrow}c}
-l_1 & (0,0)\\[2em]
\displaystyle\frac{\partial}{\partial \phi} & (1,0)\\[2em]
-l_2 & (P,s)\\[2em]
\displaystyle\frac{\partial}{\partial \psi} & (F,q)
\end{array}
\nonumber
\]
\end{minipage}%
~~~~~\begin{minipage}[t]{0.45\linewidth}
\vspace{0pt}
\includegraphics[scale=0.7]{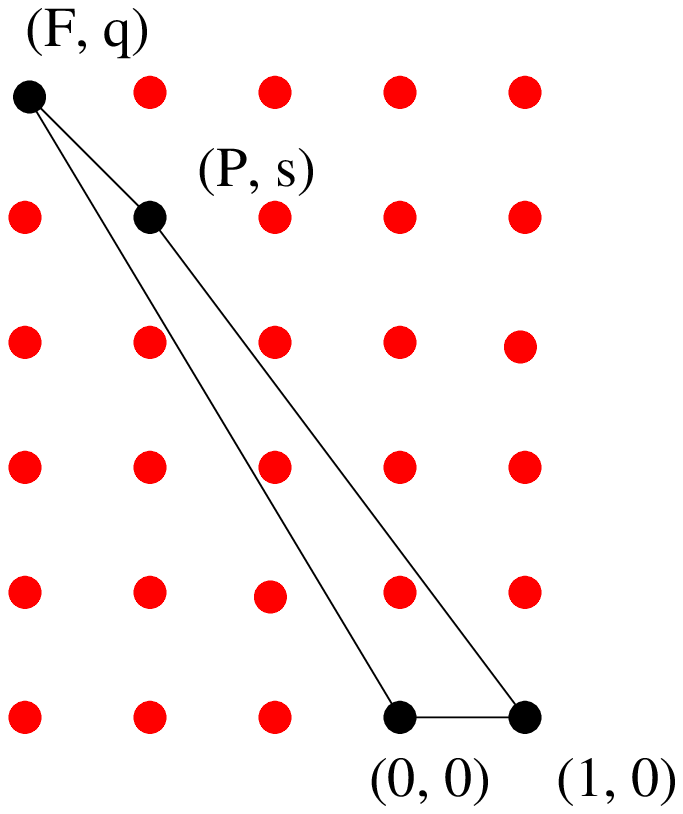}
\end{minipage}
\end{minipage}
\caption{The toric diagram for $L^{p,q,r}$. $F$ is typically negative 
and it will be renamed $-k$ in the following. The toric diagram
is given by the points $(0,0),(1,0),(P,s),(-k,q))$. The explicit example
in the Figure is $L^{1,5;2,4}$.}\label{toricdia}
\end{figure}  
where 
$F$ and $P$ correspond to any solution of (\ref{eq2}). In Figure \ref{toricdia}
we present the case $ L^{p,q;r,s}=L^{1,5;2,4}$ with $F=-3$ and $P=-2$.

Notice that in the case $r=s=\bar p$, when one gets the known case $Y^{\bar p, \bar q}$, (\ref{eq2}) has always the solution $P=0$, $F=-1$, and the toric diagram 
$(0,0)$, $(1,0)$, $(0,\bar p)$, $(-1,\bar p+\bar q)$, after a translation of $(-1,0)$ and the $SL(2,\mathbb{Z})$ transformation $((-1,0),(-\bar p,-1))$, can always be reduced to the standard form: $(0,0)$, $(1,0)$, $(2,\bar p -\bar q)$, $(1,\bar p)$.

\begin{figure}[h!!!]
\begin{minipage}[t]{\linewidth}
\begin{minipage}[t]{0.45\linewidth}
\vspace{0pt} \verb| | \\
\[
\begin{array}{c@{\rightarrow}c}
\displaystyle\frac{\partial}{\partial \phi} & (0,0)\\[2em]
-l_2 & (1,0)\\[2em]
\displaystyle\frac{\partial}{\partial \psi} & (A,p)\\[2em]
-l_1 & (X,s)
\end{array}
\nonumber
\]
\end{minipage}%
~~~~~\begin{minipage}[t]{0.45\linewidth}
\vspace{0pt}
\includegraphics[scale=0.7]{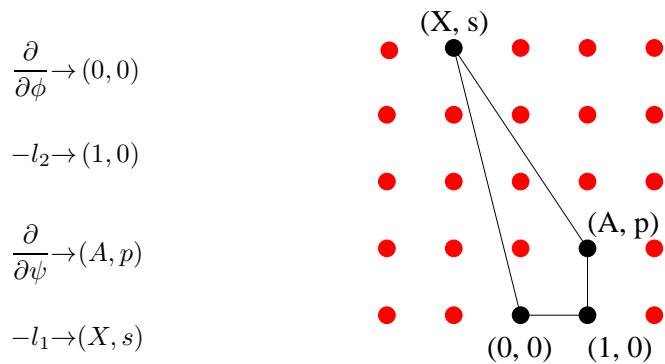}
\end{minipage} 
\end{minipage}
\caption{An alternative description for the toric diagram of $L^{p,q,r}$.
The explicit example is still $L^{1,5;2,4}$. $A$ and $X$ satisfy the linear Diophantine equation $s\,A-p\,X-q=0$}\label{to2}
\end{figure}

We have made arbitrary choices in the previous analysis: for example one could have set the vectors corresponding to $\frac{\partial}{\partial \phi}$ and $-l_2$ equal to 
$(0,0)$ and $(1,0)$ respectively, obtaining the result in Figure \ref{to2}.
Again in the Figure we present the case $L^{1,5;2,4}$; obviously the toric diagram is equivalent to the previous one.

There are other non ``minimal'' smooth CY cones over the spaces $L^{p,q;r,s}$, that correspond to $|n|>1$. Notice that all these smooth cases can be obtained by the ``minimal'' ones by multiplying the ordinates of the vertices of the toric diagram by a factor $n$: $q \rightarrow n \,q$, $s \rightarrow n \, s$. 
But now the request that the couples of integers $p,q$ and $r,s$ are pairwise coprime is no more sufficient to get a smooth manifold for a fixed $n$: one should check that the sides of the toric diagram do not pass through integers points. For example multiplying by $|n|>1$ the toric diagram in Figure 2 one never gets a smooth manifold. Instead if we multiply the diagram in Figure 1 by $ n \equiv 1 $ or $n\equiv 2$ $({\rm mod} 3)$ we find a smooth manifold (but not if n is divisible by 3). Notice also that starting from $SL(3,\mathbb{Z})$ equivalent toric diagram and multiplying by $n$ one can get inequivalent diagrams, as in our example (see also equation (\ref{class})).
These non ``minimal'' cones are $Z_n$ quotients of the ``minimal'' ones. 
We can also abandon  the request that $p,q,r,s$ are coprime and rescale them by $n$ in the case of $Z_n$ quotients, so that we can continue to use the equations $G=q$ and $Q=s$ in any case. This will be understood in the following.

To finish this discussion, we notice that in describing the gauge theory we can always choose $r \leq  s$, at least in smooth cases.
 If not so, we can exchange $r$ with $s$: this does not change
 the geometry and therefore does not change the gauge theory. In fact it is easy to check that the toric diagrams for
 $L^{p,q;r,s}$ and $L^{p,q;s,r}$ are the same up to an $SL(3,\mathbb{Z})$ transformation: start with the points $(0,0)$, $(1,0)$, $(P,s)$, $(F,q)$ with $P$ and $F$ 
satisfying  $r+F\,s-P\,q=0$. Then apply the transformation: $((\bar F, x ), (q, -F))$ where we can choose the integers $\bar F$ and $x$ to be solutions of $F \bar F+q\, x=1$
 (note in fact that $hcf(F,q)=1$ in smooth cases). The resulting toric diagram is $(0,0)$, $(1,0)$, $(\bar P, r)$, $(\bar F, q)$ with $\bar P$ and $\bar F$ satisfying:
 $s+\bar F\,r-\bar P\,q=0$. We have therefore exchanged the role of $r$ and $s$.

Analogously, when $r \leq s$, it is easy to show that it is possible to exchange the pair $(p,q)$ with the pair $(r,s)$: we have to add $(-1,0)$ to the points of the toric
 diagram  $(0,0)$, $(1,0)$, $(\tilde P,q)$, $(\tilde F,s)$ and then apply the transformation  $((-1,\lambda),(0,1))$. One gets the diagram:  $(0,0)$, $(1,0)$, $(\tilde P,q)$, $(\tilde F,s)$, where now $\tilde F$ and $\tilde P$ satisfy the relation: $p+\tilde F\,q-\tilde P\,s=0$, thus exchanging the role of $(p,q)$ and $(r,s)$.
 Note that we have used transformations with determinant $-1$; moreover we expect the changes of coordinates realizing these symmetries to be non trivial.

In conclusion we can choose $p \leq r \leq s \leq q$ without loss of generality. This ordering will always be understood in the following.


\subsection{Volumes from toric geometry}
\label{volumtoric}
Recall that the Sasaki-Einstein metrics are associated with a specific vector 
field, the Reeb vector. Using the local form of the metric~(\ref{metrica}),
it can be identified with $\frac{\partial}{\partial\tau}$.
In \cite{MSY} it was discussed a minimization procedure for determining the
Reeb vector from the toric data. Recalling that, in the dual gauge theory,
the Reeb vector is associated with the R-symmetry, the method discovered
in \cite{MSY} can be considered as the geometric dual of the a-maximization 
\cite{intriligator}. Using this algorithm one is able to compute the Reeb
vector and the volume of a Sasaki-Einstein metric on the base of a toric CY cone using only the data of the toric diagram, 
without knowing the explicit form of the metric.

We can expand the Reeb vector of a Sasakian manifold in the basis 
of vectors $e_i$ that generate the $T^3$ effective action
\begin{equation}\label{K}
K = \sum _{i=1} ^{3} b_i e_i
\end{equation}
Since in \cite{MSY} it was shown that we can always set $b_1 = 3$, we can regard $K$ as the vector $b=(3,w,t)\in {\mathbb R}^3 $. We also need the vectors $v^a$ that define the toric fan
(or equivalently the inward normals to the faces of  the polyhedral cone $\mathcal{C}$). 
The vectors $v_a$ and $b$ give enough information 
to find the volume of $L^{p,q,r}$ spaces \cite{MSY}:
\begin{equation}\label{lpqr}
{\rm vol}_{L^{p,q,r}} = \frac{\pi ^3}{3} \sum _a \frac{(v_{a-1},v_a , v_{a+1})}{(b,v_{a-1},v_a )(b,v_a , v_{a+1})}
\end{equation}  
and the volumes of the base manifolds  $\Sigma _a$ (at $r=1$) of the four special divisors associated to the four vectors of the fan\footnote{$(x,y,z)$ is the determinant of the $3 \times 3$ matrix whose rows are $x$ ,$y$ and $z$ respectively.}:
\begin{equation}\label{sigmaa}
{\rm vol}_{\Sigma _a} = 2\pi ^2 \frac{(v_{a-1},v_a , v_{a+1})}{(b,v_{a-1},v_a )(b,v_a , v_{a+1})}
\end{equation}
where $v_1,...,v_d$ are ordered in such way that the corresponding facets are adjacent to each other, with $v_{d+1} \equiv v_1 $.

The important result of \cite{MSY} is that one can find the value of $b$ 
by extremizing the volume function:
\begin{equation}\label{zeta}
Z_{L^{p,q,r}}(b) = \frac{1}{48 \pi ^3}{\rm vol}_{L^{p,q,r}}(b)
\end{equation}
It is possible to show that there is always only one critical point of $Z(b)$ in the dual cone $\mathcal{C^*}$.

We can write the function $Z(b)$ in the case of the $L^{p,q,r}$ spaces.
The four vectors $v^a$ for the case of $L^{p,q,r}$ have been determined in the
previous subsection: 
\begin{equation}
v_1=(1,0,0),\qquad v_2=(1,1,0), \qquad v_3=(1,P,s),\qquad v_4=(1,-k,q)
\label{polygon1}
\end{equation}
where $k$ and $P$ are determined by the Diophantine equation
\begin{equation}
k s+q P =r
\label{dioph1}
\end{equation}
For a better comparison with the gauge theory Section we have renamed $F=-k$.
With these values we obtain
\begin{equation}\label{zetalpqr}
Z_{L^{p,q,r}}(b) = -\frac{P^2q^2t+Pq(3qs-qt+2kst)+s(k^2st+k(3qs-2qt+st)+q(-q+s)w)}{48t((-1+P)t-s(-3+w))(kt+qw)(P(-3q+t)+k(-3s+t)+(q-s)w)}
\end{equation}
     
The minimization of this function gives algebraic equations for 
the determination of $w$ and $t$. We were not able to solve analytically
these equations. However it is easy to build a program in Mathematica
for computing numerically the value of $b$ and, consequently, the volume of the
manifolds and of the special divisors for all values of $p,q,r$. This is the
method we have used to compare the results of the volumes with the results
we will obtain on the quantum field theory side.
 
However, we can also write explicit simple expressions for the volume of
the divisors by recalling that we have a different analytic expression
for the Reeb vector. The latter is indeed given by 
$ K = \frac{\partial}{\partial \tau}$. We already provided 
the expression of $e_i$ in function of $\frac{\partial}{\partial \phi}$, $\frac{\partial}{\partial \psi}$ and $\frac{\partial}{\partial \tau}$.
From equation (\ref{K}) and the results of the previous Sections we obtain
\begin{equation}\label{kb}
b = (3, \frac{k b_1 - a_1}{c_1} , - \frac{b_1 q }{c_1} )
\end{equation}
It is easy to check numerically that this value indeed minimize the function
$Z(b)$. With the explicit form for $b$ 
we can obtain the expression of the volumes of $\Sigma _a$ as functions of 
$p$, $q$, $r$ \footnote{Remember that as explained in Section \ref{geometry}, $a_1$, $b_1$, $c_1$ are known as functions of $p$, $q$, $r$.}: 
\begin{eqnarray}\label{tetazero}
{\rm vol}_{\Sigma _{(\theta = 0,r =1)}} &=& \frac{2\pi ^2 s }{t(3s-t+Pt-sw)} = \frac{2c_1 ^2 \pi ^2 s}{b_1 q (s+b_1p)}
\nonumber\\
{\rm vol}_{\Sigma _{(\theta = \pi /2 ,r =1)}} &=& \frac{2\pi ^2 r }{(kt+qw)(3r+w(s-q)-t(k+P))}\nonumber\\ &=& \frac{2c_1 ^2 \pi ^2 r}{a_1 q (r + a_1 p)}
\nonumber\\
{\rm vol}_{\Sigma _{(x = x_1 ,r =1)}} &=& \frac{2\pi ^2 q }{t(kt+qw)} = \frac{2c_1 ^2 \pi ^2 }{ q a_1 b_1}
\nonumber\\
{\rm vol}_{\Sigma _{(x = x_2 ,r =1)}} &=& \frac{2\pi ^2 p }{(s(3-w)+t(P-1))(3r + w(s-q) - t(P+k))}\nonumber\\
 &=& \frac{2c_1 ^2 \pi ^2 p}{(r + a_1 p)(s + b_1 p)}
\end{eqnarray}

One can also find the total volume of the $L^{p,q,r}$ in function of only $p$, $q$, $r$ from the relation \cite{MSY}:
\begin{equation}\label{lvoll}
{\rm vol}_{L^{p,q,r}} = \frac{\pi}{6}\sum _a {\rm vol}_{\Sigma _a}
\end{equation}  

\section{Engineering the gauge theory}
\label{pregauge}
In this Section we present a constructive way for associating a gauge theory
with the toric diagram of $L^{p,q,r}$.  
A crucial ingredient in the identification of
the gauge theory is the remarkable correspondence between brane tilings
and toric singularities discovered in \cite{dimers}. It is based on
a correspondence between dimer models and toric geometry \cite{kenyonvafa}. 

\begin{figure}[h!!!]
\centering
\includegraphics[scale=0.8]{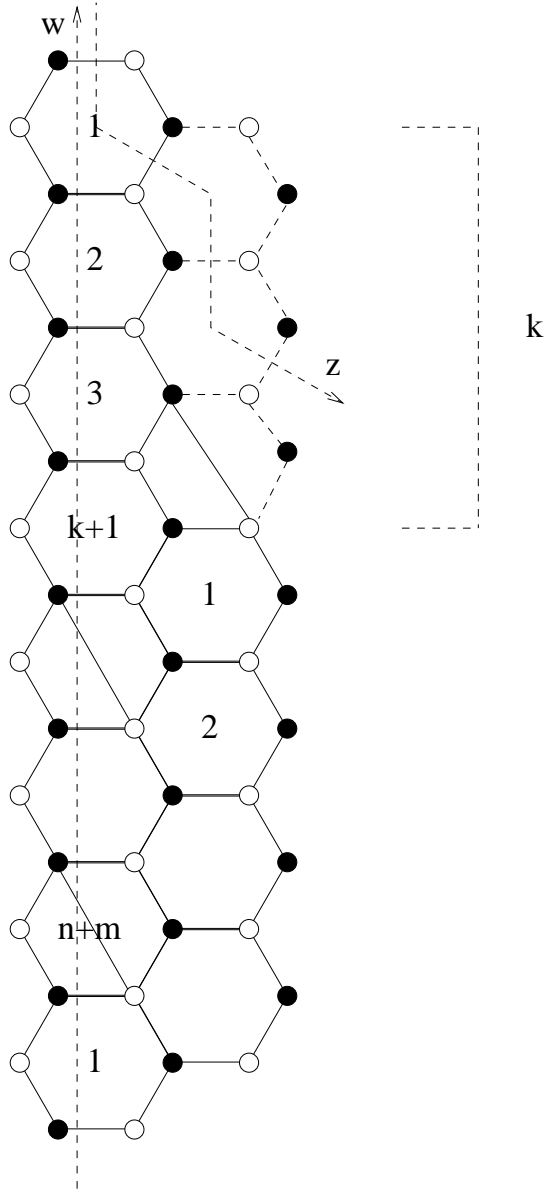}
\caption{The brane tiling corresponding to $L^{p,q;r,s}$.  The fundamental cell
is composed by $n+m$ hexagons, $m$ of which are divided in two quadrilaterals.
Each face (hexagon or quadrilateral) is a gauge group, each link a bi-fundamental
field and each vertex is a term in the superpotential given by the product
of all the fields associated with links meeting at the vertex.  
}\label{tiling}
\end{figure}

As explained in \cite{dimers}, there is an elegant way to enclose all the information needed to construct the gauge theory in a single diagram, called the periodic quiver diagram. Indeed for a gauge theory living on branes placed at the tip of toric CY cone, one can extend the quiver diagram, drawing it on a torus $ T^2$. We associate with a gauge theory a dimer model which is simply the dual of the periodic quiver and so is still defined on a torus $T^2$ \cite{dimers}; it has also a physical interpretation in terms of tilings of $D5$ and $NS5$ branes. An
example of dimer lattice is presented in Figure \ref{tiling}. Nodes are
colored in black and white and links connect only vertices of different color.
In the brane tiling, gauge groups are represented by the faces of the dimer, the edges correspond to bifundamental matter between two gauge groups, and each term in the superpotential corresponds to a vertex \cite{dimers}. 

The relation between dimers and toric geometry has been introduced in \cite{kenyonvafa} and passes through the Kasteleyn matrix, a kind of weighted adjacency matrix of the dimer. We will explain the construction of the Kasteleyn matrix
in the specific example of the $L^{p,q,r}$ singularities, 
even though the algorithm is completely general for dimers 
diagrams \cite{dimers}.
The matrix allows to associate a toric diagram with each dimer
model, and therefore with each gauge theory. The difficult part of the game is, 
given the toric singularity, to determine the brane tiling that is 
associated with it. We have determined the tiling by analogy with the 
$Y^{\bar p,\bar q}$ case. In this Section we present the result in an 
operative form; more details are given in the Appendix. 

\begin{figure}[t]
\begin{minipage}[t]{\linewidth}
\begin{minipage}[t]{0.35\linewidth}
\vspace{0pt} 
\includegraphics[scale=0.7]{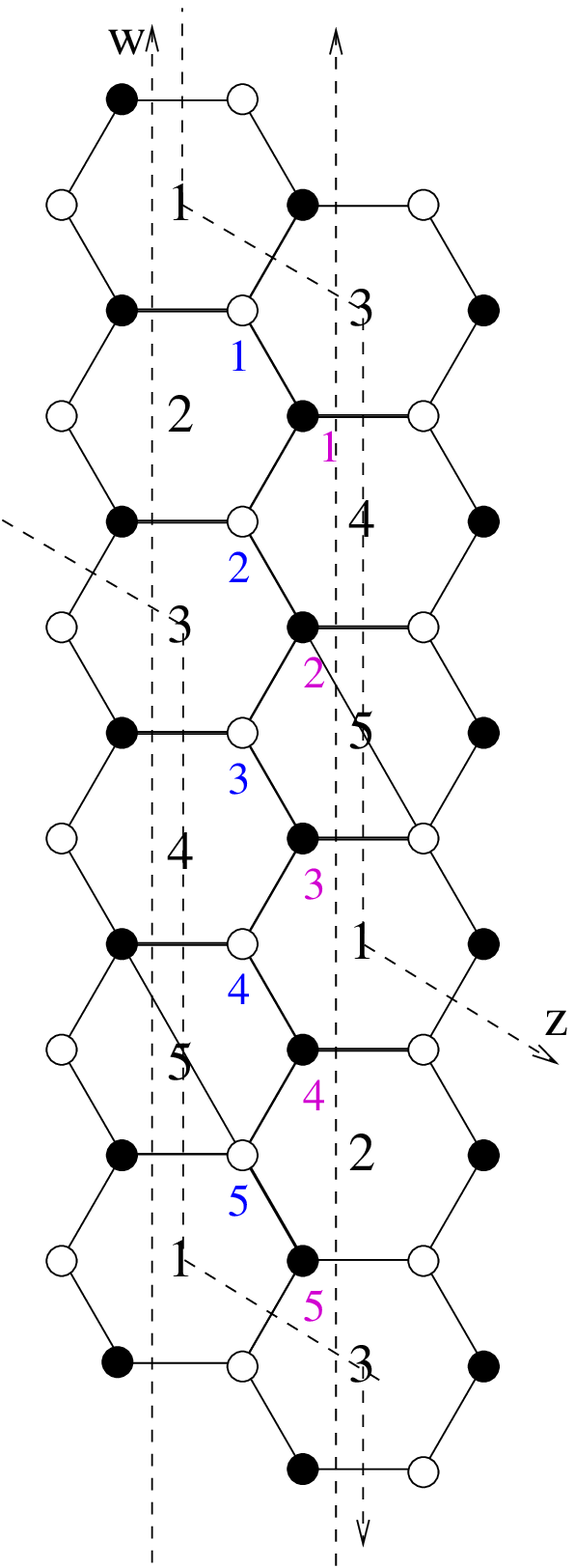}
\vspace{0pt} \hspace{-10em}
\caption{The dimer for $L^{1,5;2,4}$. In this case $n=4$, $m=1$ and $k=3$.}
\label{kast1524}
\end{minipage} 
\begin{minipage}[t]{0.5\linewidth}
\vspace{0pt} 
\centering
\[
K= \left(
\begin{array}{c|ccc@{\hspace{2em}}c@{\hspace{3em}}c}
   & 1 & 2 & 3 & 4 & 5  \\[0.5em] \hline 
 1 & 1 & 0 & 0 & w & z  \\[1em]
 2 & 1 & 1 & 0 & 0 & w  \\[1em]
 3 & wz^{-1} & 1 & 1 & 0 & 0  \\[1em]
 4 & 0 & wz^{-1} & 1 & 1 & 0  \\[1em]
 5 & 0 & -wz^{-1} & wz^{-1} & 1 & 1  \\[0.5em]
\end{array}
\right)
\]

\vspace{2.4em}

~~~~~~~~~~~~~~~\begin{minipage}[t]{\linewidth}
\begin{center}
\includegraphics[scale=0.6]{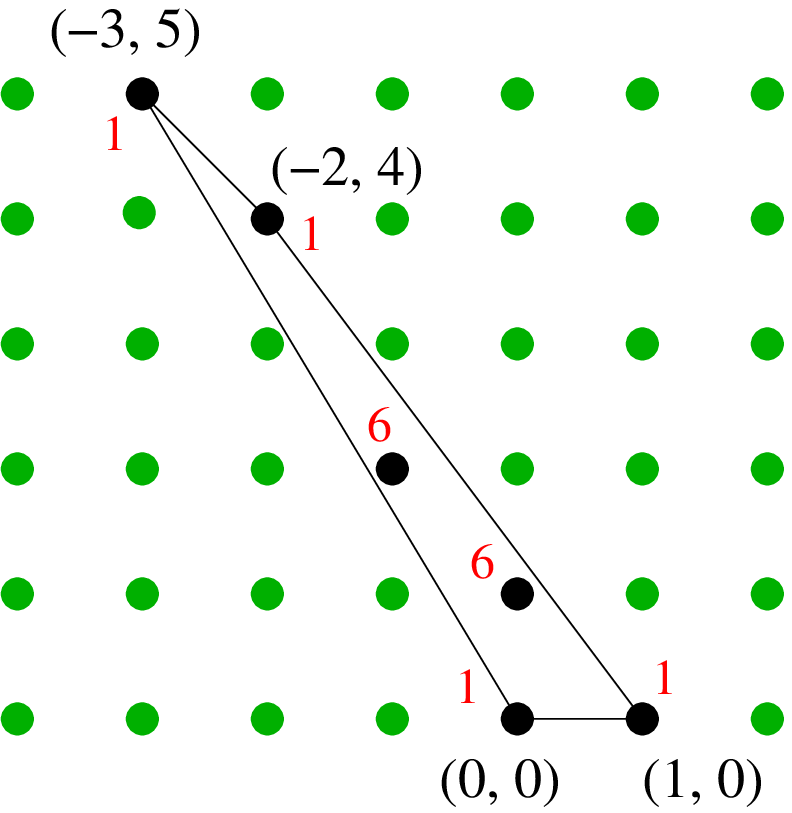}
\caption{The toric diagram for $L^{1,5;2,4}$.}
\label{torickast1524}
\end{center}
\end{minipage}
\end{minipage}
\end{minipage}
\end{figure}

We construct now a tiling sufficiently general to contain all the smooth $L^{p,q,r}$
cases, as well as many other non-smooth cases. 
We can take a (dimer) lattice 
with a fundamental cell consisting of $n$ hexagonal polygons and
$2 m$ quadrilaterals. It is convenient to picture the fundamental cell as
a sequence of $n+m$ hexagons where $m$ of them have been divided 
in two quadrilaterals as shown in Figure \ref{tiling}.  
The tiling is then obtained by covering the whole plane
with the fundamental cell repeated infinite times. 
Hexagons are identified in vertical with period $n+m$. 
The horizontal identification is more complicated: each hexagon is
identified with the hexagon in the column on the right which is shifted
down by $k$ positions, as indicated in Figure \ref{tiling}.
Now we can construct the Kasteleyn matrix $K$. The rows  
are indexed by the $n+m$ white vertices  and the columns are labeled by the $n+m$ black vertices. We have to draw two closed primitive cycles $\gamma_z$ and $\gamma_w$ on the dimer; we have chosen the cycles indicated with the dashed lines labeled with "z" and "w" in Figure \ref{tiling}. 
Then for every link in the dimer we have to add a weight in the corresponding position of $K$. The weight is equal to one if the link does not intersect the cycles $\gamma_z$ and $\gamma_w$, and it 
is equal to $z$ or $1/z$ if it intersects the cycle $\gamma_z$ according to the orientation (we choose $z$ if the white node of the link is on the right of the oriented cycle), and analogously for the $\gamma_w$ cycle. We have also to multiply all the weights for diagonal links dividing the $m$ cut hexagons by $-1$: this choice fits with the request that the product of link's signs is $1$ for hexagons and $-1$ for quadrilaterals \cite{dimers}. The explicit example for $L^{1,5;2,4}$ is shown in Figures 
\ref{kast1524} and \ref{torickast1524}.
With an appropriate ordering of the nodes, the $n+m$ by $n+m$
Kasteleyn matrix has a simple general form:
\begin{itemize}
\item{Put 1 on the diagonal.}
\item{Put 1 in the line below the diagonal and $z$ in the position $(1,n+m)$.}
\item{Put $w$ in the diagonal line that is shifted by $k$ position 
above the diagonal (beginning at $(1,k+1)$); 
whenever $w$ is shifted beyond the last column, it reappears in the equivalent column  mod $n+m$ in the form $w z^{-1}$.}
\item{Add a factor of $-w$ in the diagonal line that is shifted by $k-1$ positions above the diagonal in all the rows labeled with the same number of a divided
hexagon; whenever in this process the column is shifted beyond the last
one, the factor reappears as $-wz^{-1}$ on the left.}
\end{itemize}
  
The determinant of the Kasteleyn matrix is a Laurent polynomial $P(z,w)=\textrm{det} \,K$ called the characteristic polynomial of the dimer model. Different choices of closed primitive cycles $\gamma_z$ and $\gamma_w$ multiply the characteristic polynomial by an overall power $z^i \, w^j$.  In \cite{dimers} it was conjectured that the Newton polygon of $P$ is the toric diagram of the dual 
geometry. Recall that the Newton polygon of $P$ 
is a convex polygon in the plane $\mathbb{Z}^2$ generated by the set of integer exponents of monomials in $P(z,w)$.

We have to reproduce the toric diagram of $L^{p,q;r,s}$.
As we saw in the previous Section, the toric diagram of $L^{p,q;r,s}$ with $q\ge p$ is given by the convex polygon with vertices
\begin{equation}
[0,0],\qquad\qquad [1,0], \qquad\qquad [P,s],\qquad\qquad [F,q]
\label{polygon}
\end{equation}
where $F$ and $P$ are determined by the Diophantine equation
\begin{equation}
-F s+q P =r
\label{dioph}
\end{equation}
As explained in Section \ref{toric}, 
we can also assume $p \leq r \leq s \leq q$ without any loss of generality.
It is not difficult to see that the toric diagram derived from
the Kasteleyn matrix in the case of the proposed tiling has vertices
$(0,0)$, $(1,0)$, $(?,?)$, $(-k,n+m)$. The fourth vertex can be easily 
determined case by case using Mathematica. We now propose the following
correspondence between smooth $L^{p,q,r}$ and tiling; we have checked it
up to reasonable values of $p+q$:

\begin{itemize}
\item{For all $p,q;r,s$ with $p \leq r \leq s \leq q$ associated with the toric
diagram~(\ref{polygon}) consider a tiling with $n+m\equiv q$ hexagons of which
$m\equiv p$ have been divided. The shift is identified with $k=-F$. The
number $P$ is associated with the positions of the divided hexagons.
In the class of tiling defined by the numbers $n,m,k$, and distinguished by
the position of the divided hexagons, there is always
at least one corresponding to the diagram~(\ref{polygon}). It is not difficult to build an algorithm in Mathematica to check case by case that the proposed dimer theory correctly reproduces the geometry of $L^{p,q;r,s}$.}
\end{itemize}
The correspondence between $p,q;r,s$ and $n,m,k$ is not injective nor
surjective. Many tilings with the same values of $n,m,k$ but with
different order for the divided hexagons correspond to the
same singularity: this is reflected in the fact that the corresponding
gauge theories are connected by Seiberg dualities. Many other tilings
correspond to singular horizons, having toric diagrams with integer points
lying on the external sides. The zoo of cones with additional singularities
is quite vast: there are orbifolds (three edges), singular models 
with four sides ($L^{1,2;1,2}$ is the suspended pinch point \footnote{The class
of models with additional singularities $L^{p,q;p,q}$ has been studied in detail
in \cite{kru2}.}) and so on.
There are also possibly pathological models with multiplicity greater than
one on the vertices of the polygon. For smooth 
$L^{p,q;r,s}$ we can always find a triple $(n,m,k)$ corresponding to its
toric diagram. For small values of $p+q$ ($p+q\le 20$) and for all pairs $(p,q)$ consisting of coprime numbers, we can always put all divided hexagons 
consecutively, let us say from position $n+1$ to position $n+m$. The first
non-trivial example of two different singularities associated with the same
numbers $n,m,k$ appears for $p+q=20$: $(4,16;5,15)$ and $(4,16;9,11)$
both correspond to $n=12,m=4,k=11$. The first can be realized with consecutive
divided hexagons, the second one with divided hexagons, let us say, in positions $(1,2,3,8)$. Of course, many other choice of positions for the divided
hexagons are equivalent to the given ones.
 
Note that the Diophantine equation determine $k$ up to integer multiple of $q=m+n$, and this fits the fact that $k$ has a period of $n+m$ cells. 
Remember also that these formulae are valid for $p\leq r\leq s\leq q$. In other cases, as already explained, one can exchange $r$ with $s$ and the pair $(p,q)$ with $(r,s)$.

\section{The gauge theory}
\label{gauge}
In this Section we present the conformal gauge theories associated with
the manifolds $L^{p,q;r,s}$, we compute the central charge and the
dimensions of the fields and we compare with the geometrical results
obtained in Section \ref{toric}. 

As we saw in the previous Section, the dimer model $(n,m,k)$ that corresponds 
to the gauge theory dual to smooth $L^{p,q,r,s}$ is given by
a tiling like that in Figure \ref{tiling} with numbers:
\begin{equation}
\begin{array}{c}
\left\{
\begin{array}{l}
n\,=\,q-p\\
m\,=\,p
\end{array}
\right.
\\[1.5em]
r-k\,s-P\,q=0
\end{array}
\end{equation}
There is always a choice of positions for the $m$ divided hexagons
corresponding to this singularity. As discussed previously, in 
many simple cases the hexagons can be disposed consecutively, 
let us say from the $n+1$ to the $n+m$ position. 

As in previous Sections, we consider, without loss of generality, $p \leq r \leq s \leq q$. 

\begin{figure}[t]
\centering
\includegraphics[scale=0.7]{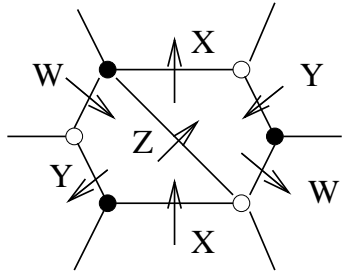}
\caption{The arrows indicate the chirality of the bi-fundamental
fields associated with the edges in the brane tiling.}\label{arrowsfig}
\end{figure}

The gauge theory on the
worldvolume of $N$ D3-branes living at the toric singularity is then
constructed with the following rules. The brane tiling gives an Hanany-Witten
construction of the gauge theory using D5 and NS5 branes \cite{dimers}.
The faces in the  brane tiling represent a finitely extended D5 brane and the edges represent
NS branes bounding the D5 branes. The configuration is roughly obtained
from the original D3 branes by applying two T-dualities which convert D3
branes in D5 branes and encode in the NS branes the geometrical information
about the singularity. Starting with $N$ D3 branes,
we obtain $N$ D5 branes covering the torus. 
Each face is then associated with a $U(N)$ gauge group.
Each edge between adjacent faces gives rise 
to a chiral field transforming
in the bi-fundamental representation of the corresponding gauge groups.
The chirality of the matter fields is determined by the natural orientation
of the edges induced by the dimer lattice. For practical purposes,
the orientation of the fields is indicated with arrows in Figure \ref{arrowsfig}.
\begin{figure}[t]
\centering
\includegraphics[scale=0.7]{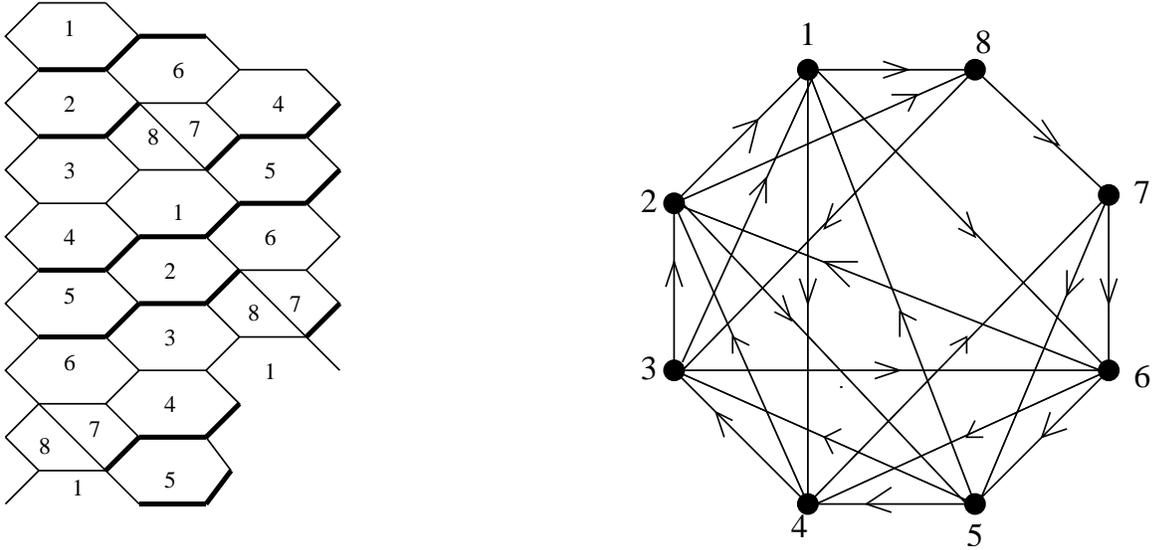}
\caption{The brane tiling for $L^{1,7;3,5}$ and the more conventional
quiver diagram (where nodes are gauge groups and oriented links chiral
bi-fundamentals). The bold edges indicate, among the $X$ and $W$ fields,
those called $\tilde X$ and $\tilde W$, with different R charge.
In this case, $n=6$, $m=1$ and $k=2$.}\label{quiv} 
\end{figure}%
Fields associated with horizontal edges are conventionally named $X$,
those associated with diagonal edges are named $Y$ and $W$ depending
on the slope of the edge and, finally, a special name $Z$ is
reserved to fields connecting quadrilaterals. Each vertex in the tiling
corresponds to a superpotential term in the gauge theory obtained by
multiplying the fields associated with the edges that meet in the given vertex.
The total number of gauge group is $n_{{\rm groups}}=p+q$, the number
of fields is $n_{{\rm fields}}= 3 q+p$ and the number of 
superpotential terms is
$n_{{\rm pot}}=2 q$. We obviously have \footnote{This relation
is a consequence of the topology of the lattice \cite{dimers}, as discussed in the Appendix.}
\begin{equation}
n_{{\rm groups}}-n_{{\rm fields}}+n_{{\rm pot}}=0
\label{tri}
\end{equation} 
With this simple set of rules we can associate a gauge theory with each
tiling. The case of $L^{1,7:3,5}$ is pictured in Figure \ref{quiv} 
with the more familiar quiver description aside.

The AdS/CFT correspondence predicts that the gauge theory dual to
$AdS_5\times L^{p,q;r,s}$ is superconformal. We can prove that the theory 
described in the previous paragraph flows in the infrared to a superconformal 
fixed point with standard methods \cite{LS}. Conformal invariance requires 
two sets of conditions. The first one is that the beta function of all the 
gauge groups vanishes. We know that, for a supersymmetric gauge theory,
the beta function is proportional to \cite{shifman}
\begin{equation}
3 T({\rm adj})-\sum_i T(r_i)(1-2\gamma_i)
\label{NSVZ}
\end{equation}
where $\gamma_i$ is the anomalous dimension of the $i$-th chiral multiplet
transforming in the $r_i$ representation of the gauge group. $T(r_i)$ is the Casimir of the $r_i$ 
representation normalized such that $T({\rm adj})=N$ and 
$T({\rm fundamental})=1/2$. Recall now that, at a superconformal fixed point, 
dimension and R-charge of a chiral field are related 
by $R_i=2 \Delta_i/3=2(1+\gamma_i)/3$.
In our case, where all the fields are bi-fundamentals, the vanishing
of the beta functions gives a set of $p+q$ equations,
\begin{equation}
\beta_k=2+\sum_{i_k}(R_{i_k}-1)=0
\label{beta}
\end{equation}
where the sum is extended only to the fields charged under the $k$-th group.
Conditions~(\ref{beta}) can be re-interpreted as the requirement
that the R-symmetry is anomalous free \footnote{Recall that the charge
of the fermion in the chiral multiplet is $R-1$. The charge
of a gaugino is conventionally normalized to $1$.}. 
This is not a mysterious result.
As is well known, in a superconformal theory, the trace of the stress-energy
tensor and the $U(1)_R$ anomaly belong to the same supermultiplet.
As a consequence, the vanishing of the beta function gives the same condition
as the cancellation of the anomaly.

The second sets of conditions come from the requirement that each term
in the superpotential has dimension three, which means R-charge 2.
We have a total of $2 q$ such conditions. We see from relation~(\ref{tri}) that the number of constraints imposed by conformal invariance 
is equal to the number of unknown R-charges. However, not all the
conditions are linearly independent. It is not difficult to see that,
for integers $p,q,r,s$ corresponding to smooth metrics $L^{p,q;r,s}$,
there are exactly three redundant conditions. This means that all the
R-charges can be expressed in terms of three 
unknown variables. The reason why we found a three parameter solution of the 
set of conditions for conformal invariance is that, given a non-anomalous
R-symmetry, we can obtain another one by adding all non-anomalous 
global $U(1)$ symmetries. In our specific case, we can see that there
are exactly three non-anomalous global $U(1)$ symmetries from the 
number of massless vectors in the $AdS$ dual.
Since the manifold is toric, the metric has three $U(1)$ isometries.
One of these (the Reeb one) corresponds to the R-symmetry while the other two
are related to non-anomalous global $U(1)$s. A third gauge field in $AdS$ comes
from the reduction of the RR four form on the non-trivial $S^3$
in $L^{p,q;r,s}$; in the supergravity literature the corresponding
vector multiplet is known as the Betti multiplet. 

The existence of a three parameter solution of the conditions for conformal
invariance is enough to establish the existence of
a fixed point. The gauge theory indeed depends on $n_{{\rm group}}$ gauge
couplings and $n_{{\rm pot}}$ superpotential couplings. 
Conformal invariance only impose $n_{{\rm group}}+n_{{\rm pot}}-3$ conditions
on these couplings, generically leaving a three-dimensional complex space
of solutions corresponding to a three dimensional manifold of conformal
field theories.

We can go further and solve explicitly the conditions for conformal invariance
in terms of three unknown R-charges, $x,y,z$. In this process, we discover that
many fields have the same R-charge. Some examples are reported
in the Appendix.  We can extrapolate from the results and give a general 
recipe: we find that, in all cases, there are
\begin{itemize}
\item{$p$ $Z$ fields with charge $z$}
\item{$r$ $X$ fields with charge $x$}
\item{$q$ $Y$ fields with charge $y$}
\item{$q-r$ $X$ fields with charge $x+z$; they will be called $\tilde X$
from now on, to distinguish them from the other $X$ fields.}
\item{$q-s$ $W$ fields with charge $2-y-x$}
\item{$s$ $W$ fields with charge $2-x-y-z$; they will be called $\tilde W$
from now on, to distinguish them from the other $W$ fields.}
\end{itemize}
The distinction among the $X$ and $W$ fields is not easy to formalize.
We will try to do that by explaining the following graphical rule:
\begin{itemize}
\item{Start at the bottom-right vertex of one divided hexagon.
Call $\tilde W$ the $W$ field attached to this vertex.
 Always by moving 
down in the vertical direction make the following steps. Move down
by $k$ hexagons. If you are at the bottom-right vertex of a
regular hexagon, call $\tilde X$ and $\tilde W$ the $X$ and $W$ fields
meeting at this ver\-tex and then move down by $k+1$ steps; instead if you are at the bottom-right vertex of
a divided hexagon call $\tilde W$ the $W$ field attached to this vertex and move down by $k$ steps.  
Repeat the procedure until you go back to the starting point.
All the fields you have encountered in this way
define the set of $\tilde X$ and $\tilde W$ fields. In Figure 3 these
fields are marked in bold.}
\end{itemize}

Note that in order to reproduce the correct counting of the different kinds of fields, during the ``cycle'' described in the algorithm above, one has to touch all the $m$ cut hexagons and $q-r$ of the normal hexagons. In this way one gets $q-r$ fields of $\tilde X$-type and $q-r+p=s$ fields of $\tilde W$-type.
The existence of a single ``cycle'' with such features is also the requirement that allows to find a correct distribution of cuts among the $n+m$ hexagons for a given
 $L^{p,q;r,s}$ theory. Further details are given in the Appendix.

One can check that this procedure works exactly in the case where
$p,q,r,s$ define smooth manifolds. 
It is not difficult to verify that for $r=s=\bar p, p=\bar p-\bar q,
q=\bar p+\bar q $ we recover the quiver gauge theory for $Y^{\bar p,\bar q}$
determined in \cite{benvenuti}. The $Y$ and $Z$ fields correspond to the fields denoted
with the same name in \cite{benvenuti}. The $\tilde W$ and $X$ fields are now
degenerate, in the sense that they connect the same pair of gauge groups,
and corresponds to the $U_\alpha$ doublets in \cite{benvenuti}. Finally
the $\tilde X$ and $W$ fields are degenerate and correspond to the
$V_\alpha$ doublets.

At the fixed point, only one of the possible non-anomalous R-symmetries enters
in the superconformal algebra. It is the one in the same multiplet as the 
stress-energy tensor. The actual value of the R-charges at the fixed point can 
be found by using the a-maximization technique \cite{intriligator}. As shown in
\cite{intriligator}, we have to maximize the a-charge \cite{anselmi}
\begin{equation}
a(R)=\frac{3}{32}(3 {\rm Tr} R^3-{\rm Tr} R)
\end{equation}
If we sum the conditions~(\ref{beta}) for all gauge groups, we discover that
\begin{equation}
0=p+q+\sum_i(R_i-1)={\rm Tr} R
\end{equation}
so that we can equivalently maximize ${\rm Tr} R^3$.

In our case,
\begin{eqnarray}
a &= \frac{9}{32}(p+q+s(1-x-y-z)^3+(q-s)(1-y-x)^3+q(y-1)^3\nonumber\\ 
&+ r(x-1)^3+(q-r)(x+z-1)^3+p(z-1)^3)
\label{acharge}
\end{eqnarray}
 
The results of the maximization give a complete information about 
the values of the central charge and the dimensions of chiral operators
at the conformal fixed point. These can be compared with the prediction
of the AdS/CFT correspondence \cite{gubser,gubserkleb}. 
The first important point is that the central charge is related to the volume 
of the internal manifold 
\cite{gubser}
\begin{equation}
a=\frac{\pi^3}{4 {\rm Vol}(L^{p,q;r,s})}
\label{central}
\end{equation}
Moreover, recall that in the AdS/CFT correspondence a special role is played
by baryons. The gravity dual describes a theory with $SU(N)$ gauge groups,
in our case $\prod_i^{p+q} SU(N)$. The fact that the groups are $SU(N)$ and
not $U(N)$ allows the existence of dibaryons. Each bi-fundamental field
$\Phi_\alpha^\beta$ gives rise to a gauge invariant baryonic operator
$$ \epsilon^{\alpha_1...\alpha_N} \Phi_{\alpha_1}^{\beta_1}...\Phi_{\alpha_N}^{\beta_N}
\epsilon_{\beta_1...\beta_N}$$
There is exactly one non-anomalous baryonic symmetry. It is sometime convenient
to think about it as a non-anomalous combination of $U(1)$ 
factors in the enlarged $\prod_i^{p+q} U(N)$ theory.
In the AdS dual the baryonic symmetry corresponds to the Betti multiplet
(the reduction of the RR four-form) and the dibaryons are described by
a D3-brane wrapped on a non-trivial three cycle.
The R-charge of the $i$-th field can be computed in terms of
the volume of the corresponding cycle $\Sigma_i$ using the formula \cite{gubserkleb}
\begin{equation}
R_i=\frac{\pi {\rm Vol}(\Sigma_i)}{3 {\rm Vol}(L^{p,q;r,s})}
\label{baryons}
\end{equation}
All the volumes appearing in these formulae can be extracted from
Section \ref{toric} and compared with the result of the a-maximization.
Although we cannot analytically solve the maximization equations
for all $p,q,r,s$, it is not difficult to compare the results
of the two Sections using Mathematica. Both the geometrical 
computation using toric geometry and the a-maximization involve
solving algebraic equations, and this can be easily automatized.
Several detailed examples of the comparison
are presented in the Appendix. The agreement of 
the two methods is absolutely remarkable. 

In Table 1 we present the charges
of the chiral fields. We also included the assignment of charges
under the baryonic $U(1)_B$ symmetry since it plays a special role among the 
global symmetries. The $U(1)_B$ can be identified 
using the following AdS-inspired arguments. 
All fields should have non zero charge since they all define corresponding 
dibaryons, described in the gravitational
dual as D3 branes charged under the RR four-form. Moreover, the
cubic anomaly for $U(1)_B$ should vanish since no Chern-Simons term
in the five dimensional supergravity can contain three vector fields
coming from reduction of the RR four-form \cite{benvenuti}.
A little patience shows that the assignments in the Table are the ones
that satisfy these conditions.
\begin{table}[h!!!!!!]
\begin{center}
\begin{tabular}{|c|c|c|c|c|c|c|} \hline  
\verb| | & \raisebox{-0.28em}{$Z$} & \raisebox{-0.28em}{$X$} & \raisebox{-0.28em}{$Y$} & \raisebox{-0.28em}{$\tilde X$} & \raisebox{-0.28em}{$W$} & 
\raisebox{-0.28em}{$\tilde W$}\\[0.5em] \hline 
\raisebox{-0.28em}{$U(1)_R$} & \raisebox{-0.28em}{$z$} & \raisebox{-0.28em}{$x$} & \raisebox{-0.28em}{$y$} & \raisebox{-0.28em}{$x+z$} & \raisebox{-0.28em}{$2-y-x$} & 
\raisebox{-0.28em}{$2-x-y-z$} \\[0.5em] \hline 
\raisebox{-0.28em}{$U(1)_B$} & \raisebox{-0.28em}{$q$} & \raisebox{-0.28em}{$-s$} & \raisebox{-0.28em}{$p$} & \raisebox{-0.28em}{$q-s$} & \raisebox{-0.28em}{$s-p$} &
\raisebox{-0.28em}{$-r$}  \\[0.5em] \hline 
\end{tabular}
\end{center}
\begin{center}
Table 1: charges under the R-symmetry and the baryonic symmetry.
\end{center}
\end{table}

\section{Adding fractional branes} 
\label{fractional}
In this Section we discuss the inclusion of fractional branes,
the duality cascade and the infrared physics of the non-conformal theories.

\begin{figure}[t]
\centering
\includegraphics[scale=0.4]{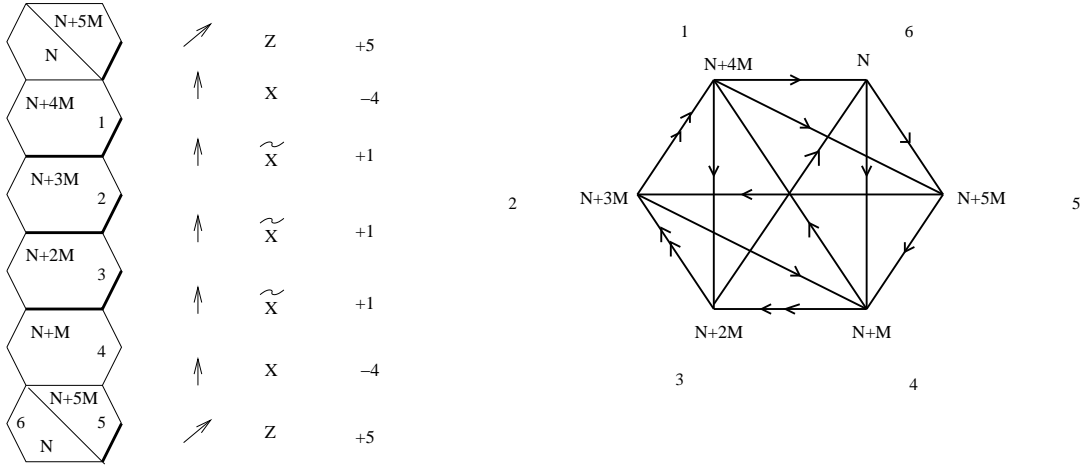}
\caption{The quiver for $N$ physical
and $M$ fractional branes in $L^{1,5;2,4}$. The number of colors
are determined using the baryonic symmetry. Double arrows indicate
that two different bi-fundamental fields are connecting two
nodes.}\label{baryo} 
\end{figure}

Since the topology of smooth $L^{p,q;r,s}$ is $S^3\times S^2$ we expect the
existence of a single type of fractional brane we can add in order
to obtain non-conformal theories. The fractional brane is a D5 wrapped
on the $S^2$ which is vanishing at the tip of the cone. The addition of
$M$ fractional branes gives a theory with the same number of gauge groups,
but with different number of colors
\begin{equation}
\prod_i^{p+q} SU(N_i)
\end{equation}
and the same bi-fundamental fields as before. The numbers $N_i$
correspond to the unique assignment of gauge groups that lead to a 
non anomalous theory. These numbers can be easily determined from 
the baryonic symmetry: the difference between the number of colors of
gauge groups connected by a bi-fundamental is $M$ times the baryonic charge
of the chiral field, with the convention that fields of opposite chirality
contribute with opposite sign. As a practical recipe, one can start
with a face in the tiling (or a node in the dual quiver), assign to it
a conventional number of colors $N$, find the numbers in the adjacent
faces using the previous rule and then continue until all numbers $N_i$
are determined. As an example, the case of $L^{1,5;2,4}$ is pictured in
Figure \ref{baryo}. The method works for the following reasons
\cite{kaza,hananyuranga}. First of
all, the cancellation of all anomalies for the baryonic symmetry implies
that the sum of charges for all the bi-fundamental fields starting from
a given face must vanish
\begin{equation}
{\rm face}\,  k:\qquad\qquad \sum_{i_k} q_{i_k} =0
\end{equation}
This condition ensures that, with the previous assignment, all gauge groups
are not anomalous. In addition, the fact that the baryonic symmetry is
a combination of the $U(1)$ factors in $\prod_i^{p+q} U(N_i)$
guarantees that the result is independent of the choice of path on the
tiling used for determining the $N_i$s.

\begin{figure}[t]
\centering
\includegraphics[scale=0.4]{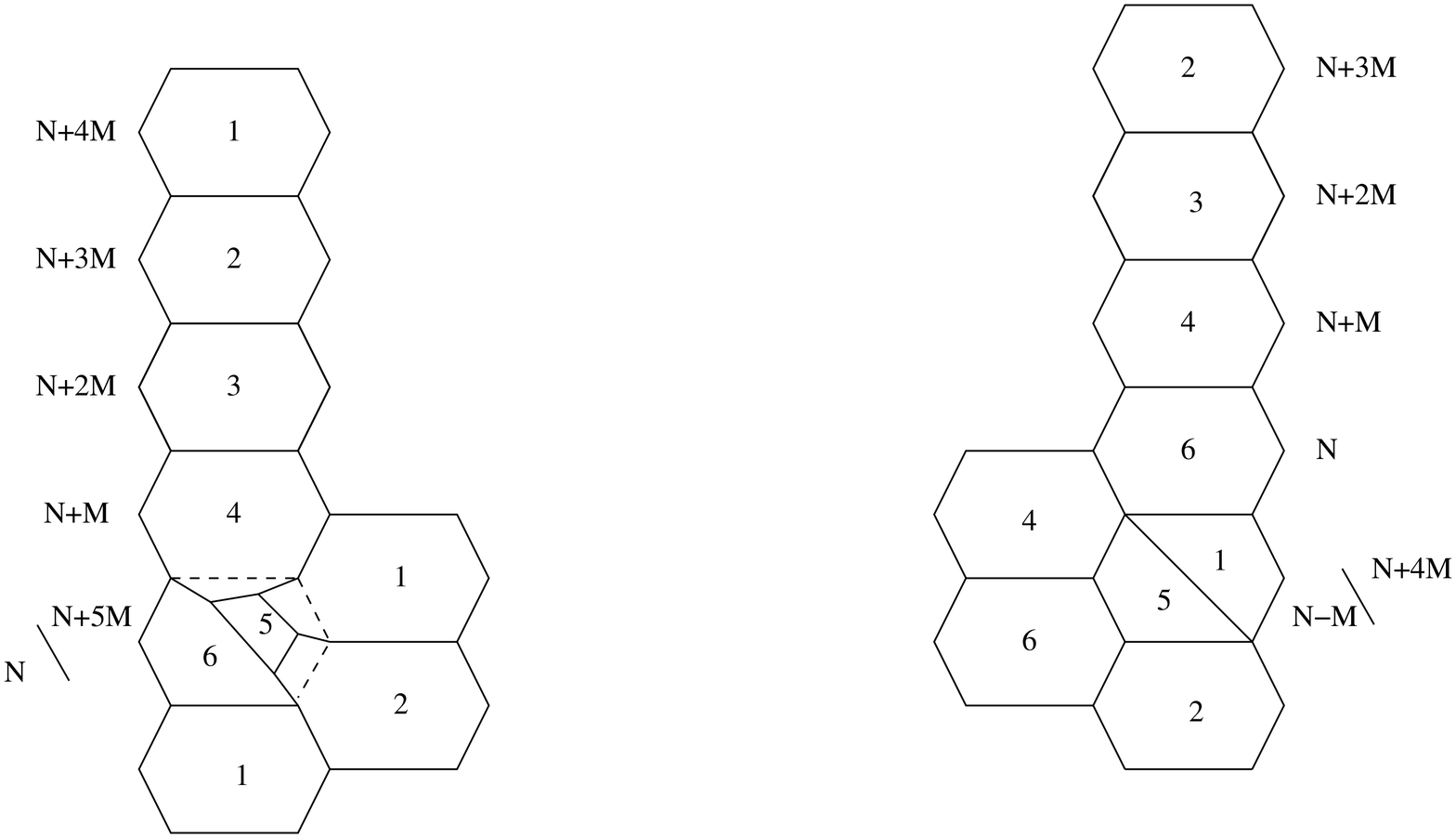}
\caption{Duality cascade for $L^{1,5;2,4}$. The fifth gauge group undergoes
a Seiberg duality. This amounts to deform it as indicated in the left 
picture. The new vertices where only two edges meet are quadratic terms
in the superpotential. The corresponding massive fields should be integrated
out. Rearranging the vertices and recalling that under Seiberg duality
$N_c\rightarrow N_f-N_c$ (the fifth gauge group will change number of colors 
from $N+5M$ to $N-M$), we get the picture on the right. 
We obtain the original tiling with a relabeling of the gauge groups.
The net effect on the theory is the shift $N\rightarrow N-M$.
We refer to \cite{dimers} for a description of Seiberg duality in
the brane tiling context.}\label{duali} 
\end{figure}

The theories with fractional branes undergoes repeated Seiberg dualities
thus giving rise to cascades. This phenomenon is familiar for the
conifold \cite{ks} and has been studied for many other quiver theories
associated with toric singularities \cite{hanany}, including the $Y^{\bar p,\bar q}$
case \cite{kleb}. One can easily verify that cascades exist also
for $L^{p,q;r,s}$. For simplicity, we consider the case of $L^{1,5;2,4}$.
As indicated in Figure \ref{baryo}, the gauge theory is
\begin{equation}
\prod_{k=0}^5 SU(N+kM)
\end{equation}
A Seiberg duality on the gauge group with the largest number of colors leads 
to a theory of the same form with $N\rightarrow N-M$. 
Perhaps, the simplest way to 
check this
is to use the graphical method discussed in \cite{dimers} (see Figure \ref{duali}).

From the point of view of the AdS/CFT correspondence, we could expect
to find a supergravity solution describing the cascade along the lines
of the Klebanov-Strassler one \cite{ks}. Certainly, it is possible
to construct a solution with fractional branes
on the singular cone \cite{gepner}. This is
 the analogous of the Klebanov-Tseytlin
solution for the conifold \cite{KT}. This kind of background has been
constructed for the $Y^{\bar p,\bar q}$ case in \cite{kleb}.
However, it is plausible that, in most cases, 
there is no regular supersymmetric
solution describing the infrared physics of the cascade. The
counterpart of this statement in quantum field theory 
is the fact that all these gauge theories have no supersymmetric vacuum.
Here we briefly discuss the motivation for these negative results 
both from the supergravity and the gauge theory side \cite{unpublished,berenstein,hananyuranga2,mavasesipuo!}. 

From the point of view of supergravity, it is known that the cones
over smooth $L^{p,q;r,s}$ manifolds have no complex deformations. The reason
is a mathematical result \cite{altmann} that states that
deformations of isolated Gorenstein toric singularities 
are in one-to-one correspondence with decomposition
of the toric diagram in Minkowski sums of polytopes \footnote{The
Minkowski sum of a convex polygon $P$ is a set of other convex polygons
$P_i$ such that all $t\in P$ can be written as $t=\sum t_i, t_i\in P_i$.}.
In the class of toric diagram which are the convex hull of four points
and corresponds to smooth horizons, only the conifold and its relatives 
can be decomposed. Clearly, the toric diagram in Figure \ref{toricdia} 
cannot be decomposed as sum of polytopes.  
This means that the cones over smooth $L^{p,q;r,s}$ have no complex deformations.
The argument is not definitive.
It rules out the possibility of imitating the Klebanov-Strassler
solution, where the original conifold singularity has been deformed.
One could argue that more general solution may exists. 
The Klebanov-Strassler metric has a metric that is conformally Calabi-Yau, 
but more general supersymmetric solutions exist. 
One could for example expect a solution based on $SU(3)$ structure
like that describing the baryonic branch of Klebanov-Strassler \cite{noi}. 
However all type IIB supersymmetric solutions with $SU(3)$ structure
must have complex internal manifolds \cite{frey,Mariana} 
and the argument in \cite{altmann} just forbids the existence
of complex deformations of our cone. It seems that supersymmetric
solutions with non complex manifolds exist in the class of $SU(2)$
structures \cite{dallagata}. In any case, if a solution exists
it is quite non trivial. 
  
This discussion is mirrored in quantum field theory by the fact that
all these theories flow in the infrared to cases where at least 
one gauge group has number of flavors less than the number of colors.
In this situation, we expect the generation of a non perturbative
ADS superpotential that will destabilize the vacuum, like in SQCD.
Let us examine the concrete case of $L^{1,5;2,4}$ for definiteness. 
Ignoring minor subtleties \footnote{At the last step of the cascade
one gauge factor has the same number of colors and flavors and
we should study 
the quantum deformed moduli space of this theory instead of naively
perform a Seiberg duality \cite{aharony,GHK}. The result however
does not change.}, the cascade will reach a point where $N=0$.
The gauge group is now 
\begin{equation}
SU(M)\times SU(2M)\times SU(3M)\times SU(4M)\times SU(5M)
\end{equation}
The $SU(5M)$ factor has only $4M$ flavors. We can make two kinds of
$SU(5M)$ gauge invariant  mesons ${\cal M}_{14}\sim YX$ and ${\cal M}_{12}\sim Y\tilde W$ and combine them in a square matrix ${\cal M}=({\cal M}_{12},{\cal M}_{14})$. A non perturbative
$1/{\rm det}({\cal M})^\alpha$ superpotential will be generated.
The presence of a tree-level superpotential does not seem to help: 
the runaway behavior induced by the ADS superpotential wins over
the stabilizing effect of the tree-level superpotential.
The only other relevant terms in the superpotential
are indeed linear both in ${\cal M}$ and in the remaining fields leading
to a superpotential that schematically reads
\begin{equation}
\frac{1}{{\rm det}({\cal M})^\alpha}+  {\cal M}_{12}\tilde X +{\cal M}_{14} W+...
\end{equation}
without any other occurrence of $W$. There is no
supersymmetric vacuum for finite values of the fields.
We have not studied the general case with arbitrary
$p,q,r,s$ but we strongly expect that the situation encountered for
$L^{1,5;2,4}$ is generic. In known examples
(reference \cite{hanany uranga} contains the discussion of 
several different class of singularities), whenever the geometry forbids
the existence of a deformation the gauge theory has a runaway 
potential. 

It is interesting to observe that our discussion is valid
for the case of smooth $L^{p,q:r,s}$. For particular values of (non coprime) 
$p,q,r,s$, the manifolds $L^{p,q:r,s}$ 
exhibit additional singularities. In some cases 
of diagrams with four vertices but with integer points on the sides,
there are complex deformations. For example, $L^{1,2:1,2}$ is called
in the literature the Suspended Pinch Point and it is known to admit
a deformation. It would be interesting to understand if our improved
knowledge of the metric for these singular horizons can help in constructing 
new regular supergravity duals of $N=1$ supersymmetric gauge theories.

\section{Conclusions}
In this paper we presented the superconformal gauge theory dual
to the type IIB background $AdS\times L^{p,q,r}$ constructed in \cite{CLPP,MSL}.
The metric has only $U(1)^3$ isometry thus making difficult to
determine the dual CFT using only symmetries. Fortunately, 
we could take advantage of an important progress recently made in
the study of the correspondence between toric singularities and CFTs,
the brane tiling construction \cite{dimers}. This ingredient,
combined with the analysis of the toric geometry of $C(L^{pqr})$,
allowed us to determine the gauge theory.

The number of explicit metrics for Sasaki-Einstein horizons than can be used
in the AdS/CFT correspondence is rapidly increasing. There are two aspects
of the correspondence where these metrics can give interesting results,
the conformal case and the non-conformal one.

On the conformal side, the most striking results come from the comparison
between conformal quantities at the fixed point and volumes. The geometrical
counterpart of the a-maximization \cite{intriligator} is the
volume minimization proposed in \cite{MSY} for determining the Reeb vector
for toric cones. The agreement of the two types of computation
suggests that some more deep connection between CFT quantities and
geometry is still to be uncovered. In particular, it would be interesting
to understand this agreement from the point of view of the brane tiling
construction. To demystify a little bit the importance of having an
explicit metric, we should note that all relevant volumes are computed
for calibrated divisors. This means that these volumes can be computed
from the toric diagram without actually knowing the metric. Explicit
formulae are given in \cite{MSY}: the volume minimization procedure
only relies on the vectors defining the toric fan. Therefore,
with a correspondence between toric diagram and gauge theories,
as that provided by the brane tiling, many checks of the AdS/CFT
correspondence can be done without an explicit knowledge of the
metric.

Where the metric is really necessary is the non-conformal side.
All these models typically admit fractional branes and 
can be used for the construction of new duals for confining
$N=1$ gauge theories. The existence of a duality cascade is a general
feature and this can be used for engineering string compactifications
with local throats, perhaps with multiple scales. We expect
to find regular supersymmetric backgrounds whenever the singularity
has a complex deformation. This is not the case for all the smooth manifolds
in the class $L^{p,q,r}$, except for the trivial case corresponding
to the conifold ($L^{1,1,1}=T^{1,1}$). From the quantum field theory
point of view, there is indeed no supersymmetric vacuum. At the end of the 
cascade, a destabilizing superpotential is generated, similar to that
of SQCD. The runaway potential forbids even non-supersymmetric vacua
\cite{hananyuranga}. Cases of singularities which admit deformations are 
known, for example the delPezzo singularities \cite{hananyuranga2}
(with at least five edges in the toric fan). Unfortunately, the metrics
for the cone in this case is unknown. It would be very interesting to see 
if one can find  new examples of deformable horizons, for example using the 
metrics for (singular) $L^{p,q;p,q}$.
 
\vskip 1truecm
 \noindent {\Large{\bf Acknowledgements}}
We would like to thank Sergio Benvenuti and Alessandro Tomasiello
for helpful discussions.
This work is supported in part by by INFN and MURST under 
contract 2001-025492, and by 
the European Commission TMR program HPRN-CT-2000-00131.

\vskip 1truecm

\noindent
{\Large{\bf Appendix}}
\renewcommand{\theequation}{A.\arabic{equation}}
\renewcommand{\thesubsection}{A.\arabic{subsection}}
\setcounter{equation}{0}\setcounter{section}{0}
\vskip 0.5truecm
\noindent
\subsection{The dimer model}
In this Section, by analogy with the $Y^{\bar p,\bar q}$ case, 
we construct the tiling configuration (dimer model) for the $L^{p,q,r}$ 
manifolds.
 
From the toric diagram that describes the geometry of the CY cone, it is possible to derive immediately some features of the gauge theory living on a stack of $N$ D3-brane at the tip of the cone. 

First of all the number of the gauge groups: as explained in the works of Hanany and collaborators, it is simply given by the double area of the toric diagram. A straightforward calculation yields: $Area=(p+q)/2$, so the number of the gauge groups is $p+q=r+s$.

The number of bifundamental chiral fields can be deduced from the $(p,q)$ web, which is obtained simply by taking the outward pointing normals to the sides of the toric
 diagram. The legs of the $(p,q)$ web are the integer vectors:
\begin{equation}
(0,-1) \qquad (s,1-P) \qquad (q-s,P-F) \qquad (-q,F)
\end{equation}  
and the formula for the number of fields is simply
\begin{equation}
n_{\rm{fields}}=\frac{1}{2} \sum_{i,j} \left| \textrm{det} \left( 
\begin{array}{cc}
p_i & q_i \\
p_j & q_j
\end{array}
 \right) \right|
\end{equation}
where $i,j$ run over the four vectors of the $(p,q)$ web. In our case the previous formula yields: $n_{\rm{fields}}=2(p+q)+|p-s|+|q-s|$.

To define a $\mathcal{N}=1$ gauge theory it is not enough to give the quiver diagram, that encodes only the information about the gauge groups and the bifundamental matter. One must also write the superpotential for chiral fields. As explained in \cite{dimers}, there is an elegant way to enclose all the information needed to construct the gauge theory in a single diagram, called the periodic quiver diagram. Indeed for a gauge theory living on branes placed at the tip of toric CY cone, one can extend the quiver diagram, drawing it on a torus $ T^2$. The vertices represent still gauge groups, the oriented links are bifundamental matter as usual, and the faces of the periodic quiver correspond to superpotential terms in the gauge theory: every superpotential term is the trace of the product of chiral fields in a face of the periodic quiver. From the Euler formula for a torus 
\begin{equation}
V-E+F=0
\end{equation} 
with V=vertices=number of gauge groups, E=edges=number of fields, and F=faces= num\-ber of terms in the superpotential,
 one can deduce that in our case $F=p+q+|p-s|+|q-s|$. 

Note that the numbers of gauge groups, superpotential terms and chiral fields just deduced are relative to a particular toric phase of the gauge theory: remember that a toric phase is characterized by a quiver where all the gauge groups have the same number of colors $N$, the number of D3-branes. There can be different toric phases for the same physical gauge theory, generally connected by Seiberg dualities and possibly characterized by different values for the numbers of fields and superpotential terms. We will exhibit a toric phase for the gauge theory on $L^{p,q,r}$ with content described by the numbers just derived.

To build the gauge theory from the toric data, one can apply the so called Inverse Algorithm \cite{Inv}, based on partial resolution of the toric singularity. Since a fast version of this algorithm is still lacking, we preferred to guess the gauge theory and then check it through the Fast Forward Algorithm \cite{dimers}. In fact to check the correctness of the gauge theory one must show that its moduli space reproduces the geometry: in the case of one brane $N=1$, it must be exactly the CY cone. 
In \cite{dimers} it was proposed a fast algorithm to compute the moduli space of the gauge theory using dimer technology; we briefly review it here.

The dimer model of the gauge theory is simply the dual of the periodic quiver and so is still defined on a torus $T^2$; it has also a physical interpretation in terms of tilings of $D5$ and $NS5$ branes, as explained in Section \ref{gauge}. However we will need only the fact that now gauge groups are represented by the faces of the dimer, the edges correspond to bifundamental matter between two gauge groups, and each term in the superpotential corresponds to a vertex. Note that the superpotential is a sum of positive and negative terms, with each field appearing exactly once in a positive term and once in a negative term. This is reflected by the fact that the dimer's vertices are colored with 
black or white and links connect only vertices of different color.
   
The moduli space of supersymmetric vacua is a toric symplectic cone whose toric diagram is easily reconstructed from the dimer model. The relation between
dimers and toric geometry have been introduced in \cite{kenyonvafa} and
passes through the Kasteleyn matrix, a kind of weighted adjacency matrix of 
the dimer.  
 
\begin{figure}[ht]
\begin{minipage}[t]{\linewidth}
\begin{minipage}[t]{0.3\linewidth}
\vspace{0pt} 
\includegraphics[scale=0.7]{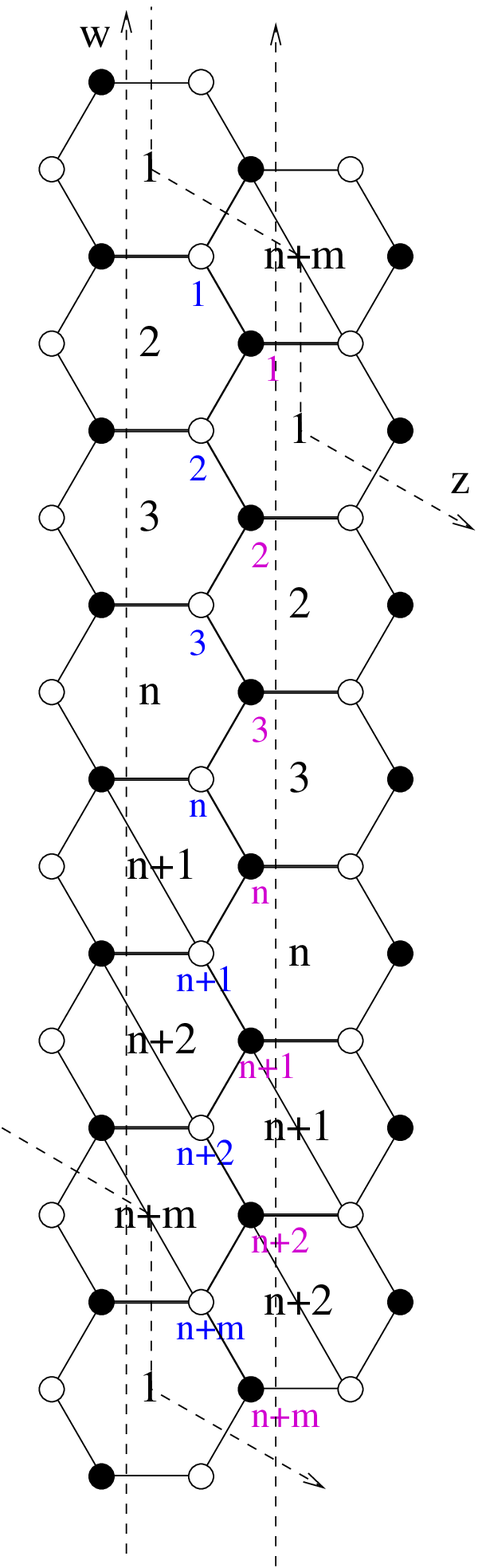}
\vspace{0pt} \hspace{-10em}
\caption{The dimer for $Y^{\bar p,\bar q}$}
\label{kast}
\end{minipage} 
\begin{minipage}[t]{0.5\linewidth}
\vspace{0pt} 
\centering
\[
K= \left(
\begin{array}{c|cc@{\hspace{2em}}c@{\hspace{2em}}cccc}
   & 1 & 2 & 3 & n & n+1 & n+2 & n+m \\[0.5em] \hline 
 1 & 1 & w & 0 & 0 &  0  &  0  &  z \\[1em]
 2 & 1 & 1 & w & 0 &  0  &  0  &  0 \\[1em]
 3 & 0 & 1 & 1 & w &  0  &  0  &  0 \\[1em]
 n & 0 & 0 & 1 & 1 &  w  &  0  &  0 \\[1em]
n+1& 0 & 0 & 0 & 1 & 1-w &  w  &  0 \\[1em]
n+2& 0 & 0 & 0 & 0 &  1  & 1-w &  w \\[1em]
n+m&wz^{-1}& 0 & 0 & 0 &  0  & 1 & 1-w  \\[0.5em]
\end{array}
\right)
\]

\vspace{2.4em}

~~~~~~~~~~~~~~~\begin{minipage}[t]{\linewidth}
\begin{center}
\includegraphics[scale=0.6]{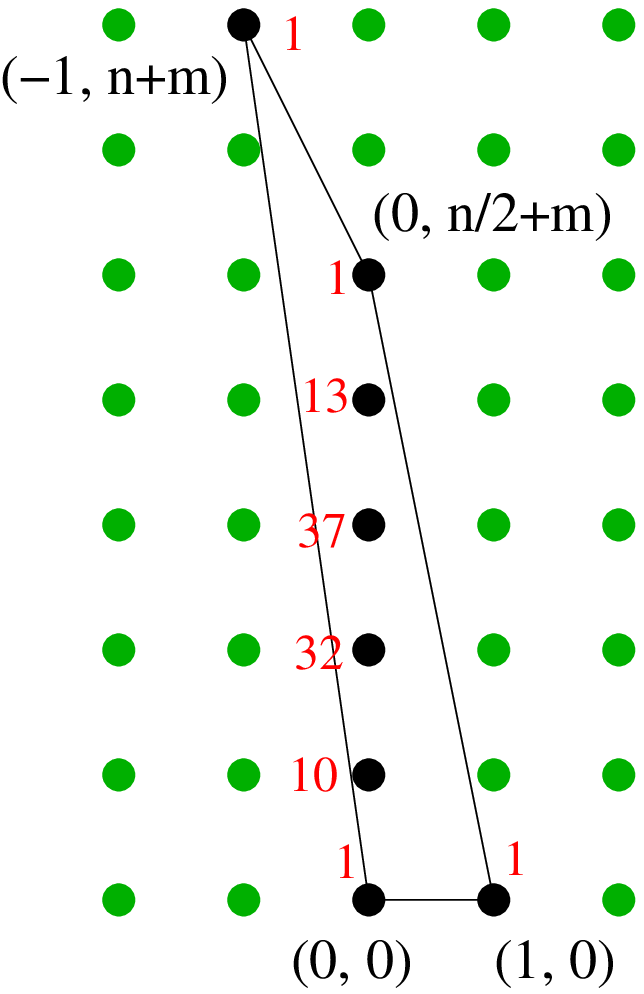}
\caption{The toric diagram for $Y^{\bar p,\bar q}$. Multiplicities refer to $Y^{5,2}$.}
\label{torickast}
\end{center}
\end{minipage}
\end{minipage}
\end{minipage}
\end{figure}

We will describe the construction of the Kasteleyn matrix on the specific example of $Y^{\bar p,\bar q}$, even though the algorithm is completely general for dimers diagrams \cite{dimers}. The gauge theories on $Y^{\bar p,\bar q}$ have dimers built only with $n$ hexagons and $2\,m$ quadrilaterals, that can be obtained by dividing in two parts $m$ hexagons drawn below the first $n$ ones (see Figure \ref{kast}). There is only one column of such hexagons and divided hexagons that are periodically identified in the vertical direction. The other $T^2$ periodicity is described by drawing other columns and identifying each face with the face in the right column shifted down by one position $(k=1)$.     

For the case of $Y^{\bar p,\bar q}$ the number $n$ and $m$ are given by: $n=2 \bar q$, $m=\bar p-\bar q$, so that we have the correct number of gauge groups: $n+2m=2\bar p$, superpotential terms: $2(n+m)=2\bar p+2\bar q$, and the correct number of fields: $3n+4m=4\bar p+2\bar q$. 

The rows in the Kasteleyn matrix $K$ are indexed by the $n+m$ white vertices (labeled with blue numbers in figure \ref{kast}) and the columns are labeled by the $n+m$ black vertices (labeled with red numbers). We have to draw two closed primitive cycles $\gamma_z$ and $\gamma_w$ on the dimer; we have chosen the cycles indicated with the dashed lines labeled with "z" and "w" in Figure \ref{kast}. 
Then for every link in the dimer we have to add a weight in the corresponding position of $K$. The weight is equal to one if the link does not intersect the cycles $\gamma_z$ and $\gamma_w$, and it 
is equal to $z$ or $1/z$ if it intersects the cycle $\gamma_z$ according to the orientation (we choose $z$ if the white node of the link is on the right of the oriented cycle), and analogously for the $\gamma_w$ cycle. We have also to multiply these weight by $-1$ for the diagonal links dividing the last $m$ hexagons in quadrilaterals: this choice fits with the request that the product of link's signs is $1$ for hexagons and $-1$ for quadrilaterals.  

The determinant of the Kasteleyn matrix is a Laurent polynomial $P(z,w)=\textrm{det} \,K$ called the characteristic polynomial of the dimer model. Different choices of closed primitive cycles $\gamma_z$ and $\gamma_w$ multiply the characteristic polynomial by an overall power $z^i \, w^j$.

The Newton polygon is a convex polygon in the plane $\mathbb{Z}^2$ generated by the set of integer exponents of monomials in $P(z,w)$. In \cite{dimers} it was conjectured that the Newton polygon of $P$ is the toric diagram 
of the geometry. This is the essential part of the Fast Forward Algorithm, since the determinant of $K$ for a given quiver allows one to reconstruct easily the dual toric geometry.
Moreover the absolute value of coefficients in $P(z,w)$ give the multiplicities of chiral fields in the Witten sigma model associated to the gauge theory.

In the case represented in Figure \ref{kast} with n=4, m=3 and k=1, corresponding to $Y^{5,2}$, the determinant of the Kasteleyn matrix is
\begin{equation}
P(z,w)=det K=1-10w+32w^2-37w^3+13w^4-w^5+\frac{w^7}{z}+z
\end{equation}
and give the toric diagram pictured in Figure \ref{torickast}. 

For general values of $m$ and even $n$ with $k=1$ it is easy to prove that the toric diagram is given by: $(0,0)$, $(1,0)$, $(0,n/2+m)$, $(-1,n+m)$, that is: $(0,0)$,
 $(1,0)$, $(0,\bar p)$, $(-1,\bar p+\bar q)$, which is the correct form given in section \ref{toric} for $Y^{\bar p,\bar q}$.

The vertex $(0,0)$ comes from the $1$ along the diagonal. The diagonal line of $1$ below the diagonal together with the element $z$ in the corner $(1,n+m)$ gives the vertex 
$(1,0)$ of the toric diagram. The diagonal line of $w$ above the diagonal, together with the element $wz^{-1}$ in the corner $(n+m,1)$ gives the vertex $(-1,n+m)$. The last point of the diagram is built with the maximal number of $w$s
(and no $z$s) in the determinant, which is $n/2+m$.

Now it is easy to generalize this to the case of arbitrary $k$. 
The simplest case is $k=0$ where $P(z,w)$ can be explicitly calculated as:
\begin{equation}
P(z,w)=(1+w)^{n+m}+(-1)^{n+m-1}z(1-w)^m \qquad k=0
\end{equation}
which gives the toric diagrams: $(0,0)$, $(1,0)$, $(1,m)$, $(0,n+m)$.
Since two sides of the diagram pass through integer points, it
corresponds to non smooth manifolds, $L^{p,q;p,q}$, already studied
in detail in \cite{kru2}.

For more general values of $k$ one gets new toric cones; in particular we looked for the values of $(n,m,k)$ that give the gauge theory for smooth $L^{p,q,r}$ and we have verified that all these smooth cases can be realized through an appropriate choice of $(n,m,k)$.   

First of all one has to understand the form of the Kasteleyn matrix in the general case where also $k$ can vary. Consider, for simplicity, the case of consecutive hexagons.
With our notations and conventions $K$ is a $(n+m,n+m)$ matrix
with the factors $1$ and factor $z$ placed in the same position as for $k=1$. But now the factors $w$, $-w$ and $wz^{-1}$ must be all shifted by $k$ positions 
along the horizontal direction: the continuous diagonal of $w$ and $wz^{-1}$ begins at position $(1,k+1)$. If a factor $w$ (or $-w$) is shifted 
beyond the last column, it reappears in the equivalent column
 mod (n+m) in the form $wz^{-1}$ (or $-wz^{-1}$). See also the explicit example of $L^{1,5,2,4}$ that corresponds to $n=4$, $m=1$, $k=3$ in figures \ref{kast1524} and \ref{torickast1524}.

So it is easy to see that the principal diagonal and the continuous diagonal below the principal one still give factors of $1$ and $z$ in the determinant, corresponding to the vertices $(0,0)$ and $(1,0)$. Moreover there's a continuous diagonal made of $k$ factors of $wz^{-1}$ and $n+m-k$ factors of $w$ corresponding to the vertex $(-k,n+m)$.
The remaining vertex is harder to find; in general the toric diagram is $(0,0)$, $(1,0)$, $(?,?)$, $(-k,m+n)$.

Now consider the case $p \leq r \leq s \leq q$. We want to fit our counting of fields, gauge groups and superpotential terms with the counting in the dimer theory 
$(n,m,k)$. Our formulae reduce to:
\begin{equation}
\begin{array}{r@{\,\,=\,\,}c@{\,\,=\,\,}l}
\# \,\textrm{gauge} & p+q & n+2\,m\\[0.5em]
\# \, \textrm{super potential} & 2\,q & 2\,(n+m)\\[0.5em]
\# \, \textrm{fields} & p+3\,q & 3\,n+4\,m\\
\end{array}
\end{equation}
  
These formulae are solved by $n=q-p$ and $m=p$. Moreover note that the toric diagram for $L^{p,q,r}$: $(0,0)$, $(1,0)$, $(P,s)$, $(F,q)$ fits with the diagram from the dimer model since $m+n=q$ and gives the identification:
\begin{equation}
F=-k
\end{equation} 
and this directly gives the way to calculate $k$ for $L^{p,q;r,s}$: $k$ and $P$ are the solution of the linear Diophantine equation: $r-k\,s-P\,q=0$.

It is easy to generalize to the case where the divided hexagons are not in consecutive
positions. The form of the Kasteleyn matrix is explained in Section \ref{pregauge}. Tilings with divided hexagons in arbitrary positions are necessary
to describe all possible $L^{p,q:r,s}$.

To summarize, the dimer model $(n,m,k)$ that defines the gauge theory dual to the geometry of smooth $L^{p,q;r,s}$ is given by:
\begin{equation}
\begin{array}{c}
\left\{
\begin{array}{l}
n\,=\,q-p\\
m\,=\,p
\end{array}
\right.
\\[1.5em]
r-k\,s-P\,q=0
\end{array}
\end{equation}
and has associated toric diagram given by: $(0,0)$, $(1,0)$, $(P,s)$, $(-k,q)$. 
Note that the Diophantine equation determine $k$ up to integer multiple of $q=m+n$, and this fits the fact that $k$ is a periodicity in a dimer with $n+m$ cells ($n$ hexagons and $m$ hexagons divided in two quadrilaterals). In our pictures we use the solution with $0\leq k \leq m+n$.

Remember that these formulae are valid for $p \leq r \leq s \leq q$. In other cases, as already explained, one can exchange $r$ with $s$ and the pair $(p,q)$ with $(r,s)$.

At this point it is not difficult to build an algorithm in Mathematica to check case by case that the proposed dimer theory correctly reproduces the geometry of $L^{p,q,r}$. We have also checked that the multiplicities of external vertices in toric diagram associated to smooth $L^{p,q,r}$ are always equal to one.

\subsection{Computing R charges}

In this subsection we present an explicit example focusing on the problem of computing the $R$ charges for the field theory, and in particular we explain the algorithm, already discussed in Section \ref{gauge}, to distinguish between fields $X$, $W$ and $\tilde X$, $\tilde W$, once a good distribution for the cuts on the hexagons has been found.

\begin{figure}[t]
\begin{minipage}[t]{0.45\linewidth}
\vspace{0pt}
\centering
\includegraphics[scale=0.7]{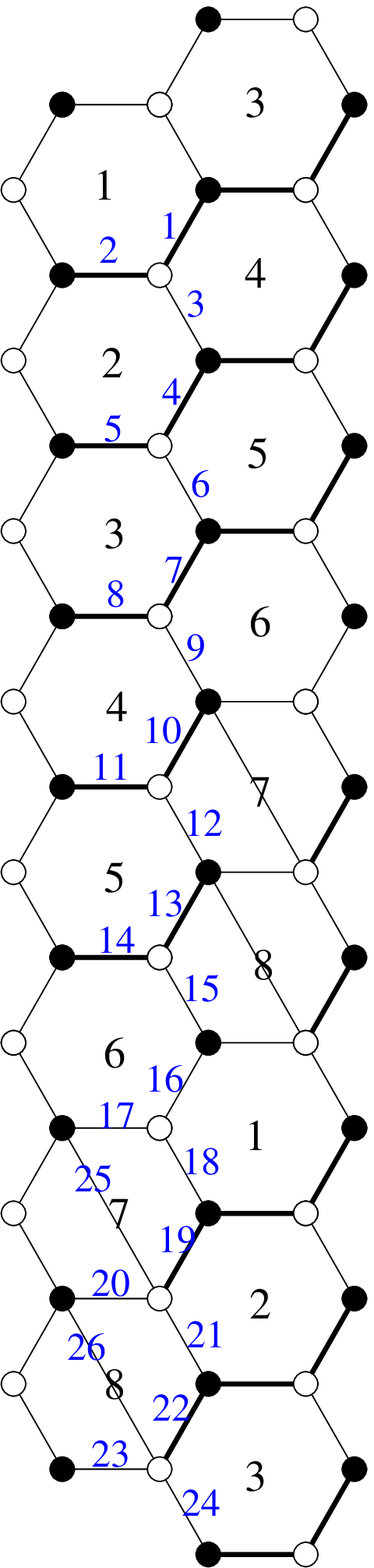}
\caption{The dimer for $L^{2,8;3,7}$}
\label{cut1}
\end{minipage}%
\begin{minipage}[t]{0.45\linewidth}
\vspace{0pt}
\centering
\includegraphics[scale=0.7]{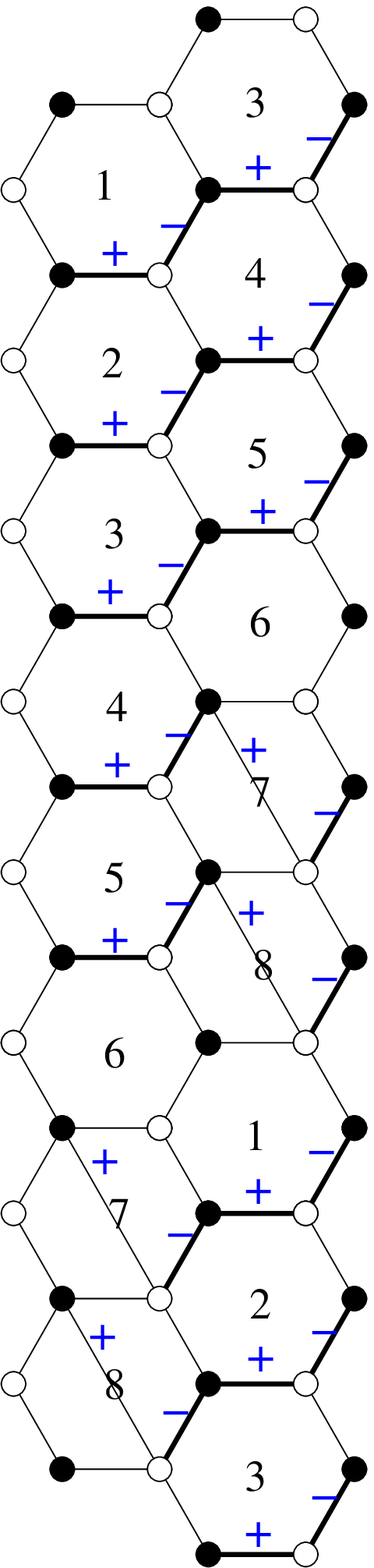}
\caption{The charge $v^1$ for $L^{2,8;3,7}$}
\label{cut3}
\end{minipage}
\end{figure}

Consider to be concrete the case $(p,q;r,s)=(2,8;3,7)$ that gives $(n,m,k)=(6,2,5)$ and $P=-4$. In this case it is possible to choose all the cut hexagons to be consecutive.  We have in total $3n+4m$ fields (=26) that we number as in Figure \ref{cut1}. The general rule is: start from the white bottom-right vertex of the first hexagon, and call the links attached to it (1,2,3): the first is a $W$-type field, the second is $X$-type field and the third is a $Y$ field. Then pass to the bottom-right vertex of the second face and continue to label the triple of fields in this order until the last white vertex is reached. At the end label the remaining $m$ $Z$ fields. So we can arrange the $3n+4m$ fields in an array divided in $n+m$ triples of fields $(W,X,Y)$ (or with tilded $W$ or $X$) and $m$ fields of $Z$ type:
\begin{small}
\[
\begin{array}{cc@{|}c@{|}c@{|}c@{|}c@{|}c@{||}c@{|}c@{||}c}
  & 1 & 2 & 3 & 4 & 5 & 6 & 7 & 8 &  \\[0.5em]
\textrm{fields}=& \left( 1,2,3 \right. & 4,5,6 & 7,8,9 & 10,11,12 & 13,14,15 & 16,17,18 & 19,20,21 & 22,23,24 & \left. 25,\, 26 \right) 
\nonumber
\end{array}
\]    
\end{small}
where the numbers in the upper line label the $n+m$ triples. Correspondingly we have an array $R$ of $3n+4m$ $R$-charges organized as in the previous formula. 

We have to find all the possible assignments of $R$ charges that make the theory conformal. The condition of zero $\beta$ function for every gauge group explained in Section \ref{gauge} becomes:
\begin{equation}
\sum_{i_{k} \in \textrm{hex}} R_{i_k}= 4 \qquad \sum_{i_{k} \in \textrm{quad}} R_{i_k}= 2
\label{faces}
\end{equation}
that is the sum of R charges of the edges of every hexagon must be equal to $4$ and must be equal to $2$ for quadrilaterals. Then we have to impose that every term in the superpotential has R charge equal to $2$; this is a condition for links attached to the same vertex:
\begin{equation}
\sum_{i_{h} \in \textrm{vertex}} R_{i_{h}} = 2.
\label{vertices}
\end{equation}

This set of $3n+4m$ linear conditions can be organized in a matricial form: $C_{ij}R_{j}=V_{i}$, where each line $i$ is  an equation that must be imposed. $C_{ij}$ is a $(3n+4m,3n+4m)$ matrix, $R_{j}$ is the array of $3n+4m$ $R$ charges and $V_{i}$ is the vector of non homogeneous known terms (made up only with $4$ and $2$).

To write the most general solution for $R_{j}$ we have to know a particular solution $v^0$ and the ker of the matrix $C_{ij}$. Then the array $R$ can be written as
\begin{equation}
R\,= \, v^0 + z\, v^1 + y\, v^2 + x\, v^3
\label{rcharge}
\end{equation} 
where $v^0$ is a particular solution and $v^1$, $v^2$ and $v^3$ are in the ker of $C$:
\begin{equation}
C_{ij}v^0_j=V_i, \qquad C_{ij}v^k_j=0, \qquad k=1,2,3.
\end{equation}
and $x$, $y$ and $z$ are arbitrary real variables.

In fact we know on physical grounds that in cases corresponding to gauge theories dual to $L^{p,q;r,s}$ the dimension of the ker of $C_{ij}$ is exactly $3$: the vectors $v^1$, $v^2$ and $v^3$ are a basis of non anomalous $U(1)$ global
symmetries  and we know that $L^{p,q;r,s}$ has two flavor $U(1)$ and one baryonic $U(1)$.
Indeed it is easy to write an algorithm to build the matrix $C_{ij}$ and check this statement in all the desired cases.

The form of $v^0$ and of two of the vectors in the kernel of $C_{ij}$ is easy to find:
\begin{small}
\[
\begin{array}{cc@{|}c@{|}c@{|}c@{|}c@{|}c@{||}c@{|}c@{||}c}
  & 1 & 2 & 3 & 4 & 5 & 6 & 7 & 8 &  \\[0.5em]
v^0 = & \left( 2,0,0 \right. & 2,0,0 & 2,0,0 & 2,0,0 & 2,0,0 & 2,0,0 & 2,0,0 & 2,0,0 & \left. 0,\, 0 \right) \\[0.6em]
v^2 = & \left( -1,0,1 \right. & -1,0,1 & -1,0,1 & -1,0,1 & -1,0,1 & -1,0,1 & -1,0,1 & -1,0,1 & 
\left. 0,\, 0 \right) \\[0.6em] 
v^3 = & \left( -1,1,0 \right. & -1,1,0 & -1,1,0 & -1,1,0 & -1,1,0 & -1,1,0 & -1,1,0 & -1,1,0 & 
\left. 0,\, 0 \right) 
\end{array}
\]    
\end{small}
and this is true for generic values of $(n,m,k)$: it is easy to show that a particular solution $v^0$ of the equations (\ref{faces}) and (\ref{vertices}) can be obtained by assigning R charge equal to $2$ to all $W$ and $\tilde W$ and zero to all the other fields.

Analogously one can get two vectors in the ker of $C_{ij}$ by assigning charge $-1$ to $W$ and $\tilde W$, $1$ to the fields $Y$, and zero to all other fields (this is the vector $v^2$) or by assigning charge $-1$ to $W$ and $\tilde W$, charge $1$ to $X$ and $\tilde X$ and zero to the other fields. It is easy to check for these assignments that the sum of charges in every vertex is zero (the superpotential is conserved by the $U(1)$ symmetry) and the sum of charges for every face is zero (cancellation of the anomaly for the $U(1)$ symmetry), thus $v^2$ and $v^3$ are in ker $C$.

The remaining $U(1)$ symmetry $v^1$ is more tricky to find; we want to show that it can be built by giving charge $1$ to all $Z$ fields, charges $(-1,0,0)$ to the $m$ triples corresponding to the cut hexagons, and charges $(-1,1,0)$ or $(0,0,0)$ to the triples corresponding to normal hexagons (in this way charge cancellation is automatically satisfied for white vertices).
Then proceed as in Figure \ref{cut3}: consider one of the divided hexagons, for instance the last (8). We have to impose charge conservation on the black top left vertex; so we must assign charge $(-1,1,0)$ to the triple corresponding to face number $5$. Now looking at the black bottom left vertex of face 5 we see that we have to put another
$(-1,1,0)$ to the triple corresponding to face $3$. Repeating this operation we find the cycle:
\begin{center}
\setlength{\unitlength}{1em} 
\begin{picture}(30,3)
\put(0,1){8} 
\put(1,1.4){\vector(1,0){2.5}} \put(4,1){5} \put(1.5,2){\small{$+k$}}
\put(5,1.4){\vector(1,0){2.5}} \put(8,1){3} \put(5,2){\small{$+k+1$}}
\put(9,1.4){\vector(1,0){2.5}} \put(12,1){1} \put(9,2){\small{$+k+1$}}
\put(13,1.4){\vector(1,0){2.5}} \put(16,1){7} \put(13,2){\small{$+k+1$}}
\put(17,1.4){\vector(1,0){2.5}} \put(20,1){4} \put(17.5,2){\small{$+k$}}
\put(21,1.4){\vector(1,0){2.5}} \put(24,1){2} \put(21,2){\small{$+k+1$}}
\put(25,1.4){\vector(1,0){2.5}} \put(28,1){8} \put(25,2){\small{$+k+1$}}
\end{picture}
\end{center}
where we have to add $k$ if we are in a cell with a cut or $k+1$ if we are in a normal hexagon (modulo $(n+m)$).
We assign $(-1,1,0)$ to all the triples corresponding to normal hexagons touched by the cycle and (0,0,0) to the remaining ones. In our concrete example we find:
\begin{small}
\[
\begin{array}{cc@{|}c@{|}c@{|}c@{|}c@{|}c@{||}c@{|}c@{||}c}
  & 1 & 2 & 3 & 4 & 5 & 6 & 7 & 8 &  \\[0.5em]
v^1 = & \left( -1,1,0 \right. & -1,1,0 & -1,1,0 & -1,1,0 & -1,1,0 & 0,0,0 & -1,0,0 & -1,0,0 & \left. 1,\, 1 \right) 
\nonumber
\end{array}
\]    
\end{small}    
All the cut hexagons are covered by this cycle; their corresponding charges are $(-1,0,0)$.
The fact that the cycle closes assures that charge is conserved also at all black vertices. Then it is easy to see that this algorithm also assures that the sum of charges for every face is zero (anomaly cancellation).

This algorithm allows to find the right disposition of fields once the distribution of cuts on the hexagons has been found. It corresponds to the algorithm on the dimer to distinguish between fields $X$, $\tilde X$ and $W$, $\tilde W$ given in Section \ref{gauge}.
It also allows to count the number of different kinds of fields: if the cycle closes after touching all the $m$ cut hexagons and $N$ normal hexagons, we must have:
\begin{equation}
m\,k+N\,(k+1)=\lambda (m+n)
\end{equation}
for some integer $\lambda$.
This is equivalent to the equation:
\begin{equation}
(k+1)\,(m+n-N)+(m+n)\,(\lambda-1)\,=\,k\,(n+2\,m)
\label{counting}
\end{equation}

Now consider the problem of finding $(p,q;r,s)$ given $(n,m,k)$: rewriting the equation $r-k\,s-P\,q=0$ we find that:
\begin{equation}
(k+1)\,r-(m+n)\,P=k\,(n+2\,m)
\label{inverse}
\end{equation}
This problem has not a unique solution for $r$ and $P$, even requiring $p \leq r \leq s \leq q$. In fact $hcf(m+n,k+1)$ is not necessarily equal to $1$.
So we see that it is possible to get the same values of $(n,m,k)$ for different theories $(p,q;r,s)$. 

When the problem (\ref{inverse}) has a unique solution, in the cases $hcf(m+n,k+1)=1$ (which is always true if $p$ and $q$ are coprime), we see by comparing expressions
 (\ref{counting}) and (\ref{inverse}) that $m+n-N=r$: they are solution of the same Diophantine equation with $0\leq (m+n-N) \, ,r \leq (m+n)$.

Using $N=m+n-r=q-r$ and equation (\ref{rcharge}) we again find that there are:
$p$ fields of $Z$ type, $r$ $X$, $q$ $Y$, $(q-r)$ $\tilde X$, $(q-s)$ $W$ and $s$ $\tilde W$.
This counting must be true also in the general case since it always gives the right expression for the central charge $a$ and for the $R$ charges of chiral fields that match the predictions of the AdS/CFT correspondence.

We conclude that the requirement for finding a good distribution of cuts for a theory dual to smooth $L^{p,q;r,s}$ is the existence of a unique ``cycle'', built with the algorithm described above, touching all the $m$ cut hexagons and $N=q-r$ of the normal hexagons. This ``cycle'' must be unique since there are only three $U(1)$ charges. 

\subsection{Some examples}

In table 2 we report the results of numerical analysis for $L^{p,q;r,s}$ with number of gauge groups $p+q\leq 16$. We list all the smooth ``minimal'' (in the sense that they are not orbifolds) cases that do not belong to the already known family $Y^{\bar p,\bar q}$. We give for every $(p,q;r,s)$ the values of $(n,m,k)$  and of the quantity $P$ entering in the toric diagram $(0\leq k \leq q-1)$, the total volume, the central charge $a$ and the R charges $R(Y)=y$, $R(\tilde W)$, $R(Z)=z$, $R(X)=x$, (the remaining charges can be computed through the expressions in function of $x$, $y$, $z$ given in the text).

The important point to note 
is that the same numbers can be obtained from the formulae in Section \ref{geometry}
and \ref{toric}.
In particular, the volume, computed through equation (\ref{volume}) or by the method described subsection (\ref{volumtoric}) following \cite{MSY}, always matches  the value for $a$ computed using the standard AdS/CFT formula. This matching is true also for the volumes of the divisors, computed as in subsection (\ref{volumtoric}), and the values of R charges of the fields $Y$, $\tilde W$, $Z$, and $X$, computed through $a$ maximization. In fact we find:
\[
\begin{array}{l@{\hspace{2em}}l}
R(Y)=\displaystyle\frac{\pi}{3}\frac{\textrm{vol}(\Sigma_{x=x_1})}{\textrm{vol}} & 
 R(\tilde W)=\displaystyle\frac{\pi}{3}\frac{\textrm{vol}(\Sigma_{\theta=0})}{\textrm{vol}} \\[1.5em]
R(Z)=\displaystyle\frac{\pi}{3}\frac{\textrm{vol}(\Sigma_{x=x_2})}{\textrm{vol}} & 
 R(X)=\displaystyle\frac{\pi}{3}\frac{\textrm{vol}(\Sigma_{\theta=\pi/2})}{\textrm{vol}}  
\end{array}
\]
The relation between the total volume and $a$ was recently demonstrated in algebraic way by \cite{kru2}, and the comparison between the geometrical formulae
for volumes of divisors and R-charges was done in \cite{tomorrow}.  

We also checked in these cases that the dimer configuration always reproduces the geometry by computing the Kasteleyn matrix. In these cases it is possible to put all
 the cuts in the $n+m$ hexagons in consecutive positions, and we believe that this is true every time there is only one choice of $(p,q;r,s)$ with
 $p\leq r \leq s \leq q$, at least for smooth cases. This happens for example if $hcf(n+m,k+1)=1$, that is for $p$ and $q$ coprime.

But as discussed few lines above the map from $(p,q;r,s)$ to $(n,m,k)$ is in general not injective; the smallest example of a pair of values of $(p,q;r,s)$ (smooth and 
 ``minimal'' case) that give the same $(n,m,k)$ is obtained for $20$ gauge groups: one can check that $(p,q;r,s)=(4,16;5,15)$ and $(p,q;r,s)=(4,16;9,11)$ using the formulas in this article give both $(n,m,k)=(12,4,11)$.
These two theories are not equivalent, but they can both be obtained with $(n,m,k)=(12,4,11)$ with appropriate choices for disposition of the cuts among the
 12+4=16 hexagons. By computation of the Kasteleyn matrix, one can see that the ``standard'' choice with all consecutive cuts corresponds to $(p,q;r,s)=(4,16;5,15)$. 
The other theory $(4,16;9,11)$ can be obtained putting the cuts in the positions $(1,2,3,8)$ where $1,2,\ldots 16$ label the $n+m$ hexagons (but there are obviously many equivalent choices obtained for example performing Seiberg dualities).
We have checked that this also happens in many other cases.

We believe that all smooth $L^{p,q;r,s}$ can be obtained by an opportune choice of $m$ cuts in the tiling with $n+m$ hexagons and shift $k$ where $(n,m,k)$ are obtained through the formulas in this article (probably this is not the only way to represent the gauge theory associated to $L^{p,q;r,s}$, but hopefully this choice can be used as a 
canonical representation). It would therefore be interesting to understand better this point and maybe write an algorithm that allows one to find the distribution of cuts in the tiling.

\vspace{0.7em}

\begin{small}
\begin{tabular}{|c|c|c|ll|llll|}
\hline
(p,q,r,s) & (n,m,k) & P & \verb| | Vol & \verb|   |a & \verb| | R(Y) & \verb| |$R(\tilde W)$ & \verb| |R(Z) & \verb| |R(X) \\ \hline
(1,5,2,4) & (4,1,3) & -2 & 5.7248 & 1.35403 & 0.632216 & 0.593238 & 0.286325 & 0.48822 \\ \hline
%
(1,7,2,6) & (6,1,5) & -4 & 4.196 & 1.84737 & 0.643002 & 0.617373 & 0.265531 & 0.474093 \\
(1,7,3,5) & (6,1,2) & -1 & 4.22672 & 1.83394 & 0.644448 & 0.594397 & 0.229036 & 0.532119 \\ \hline
%
(1,9,2,8) & (8,1,7) & -6 & 3.30707 & 2.34394 & 0.64869 & 0.629624 & 0.254868 & 0.466819 \\
(2,8,3,7) & (6,2,5) & -4 & 3.46292 & 2.23845 & 0.622014 & 0.596865 & 0.329893 & 0.451228 \\ \hline
(1,11,2,10) & (10,1,9) & -8 & 2.72765 & 2.84185 & 0.652184 & 0.637012 & 0.248404 & 0.4624 \\
(1,11,3,9) &  (10,1,4) & -3 & 2.74414 & 2.82477 &  0.653604 & 0.625732 & 0.202214 & 0.518451\\
(1,11,4,8) &  (10,1,6) & -4 & 2.75273 & 2.81596 & 0.654573& 0.615195 & 0.17831 & 0.551922\\  
(1,11,5,7) &  (10,1,7) & -4 & 2.757   & 2.81159 & 0.655113 & 0.604008 & 0.166451 & 0.574428 \\
(2,10,3,9) &  (8,2,7) & -6 & 2.8393 & 2.7301 &  0.631576 & 0.612381 & 0.31603 & 0.440013\\
(3,9,4,8) &   (6,3,5) & -4 &  2.936  & 2.64018 & 0.606051 & 0.582591 & 0.363364 &  0.447995 \\
(3,9,5,7) &   (6,3,2) & -1 &  2.95615 & 2.62218 & 0.605623 & 0.55581 & 0.347596 & 0.490971 \\
(4,8,5,7) &   (4,4,3) & -2  & 3.01039 & 2.57494 & 0.574958 & 0.545937 & 0.408526 & 0.470578\\ \hline
(1,13,2,12) & (12,1,11) & -10  & 2.32051 & 3.34045 & 0.654545 & 0.641948 & 0.244071 & 0.459436 \\
(1,13,3,11) & (12,1,5)  & -4 & 2.33298 & 3.32261 & 0.65585 & 0.633048 & 0.195902 & 0.515201 \\
(1,13,4,10) & (12,1,3)  & -2 & 2.33981 & 3.3129 & 0.656791 & 0.625167 & 0.169647 & 0.548394 \\
(1,13,5,9)  & (12,1,2)  & -1 & 2.3437 & 3.30741 & 0.657396 & 0.617305 & 0.154768 & 0.570531 \\
(1,13,6,8)  & (12,1,4)  & -2 & 2.34573 & 3.30455 & 0.657732 & 0.608744 & 0.147015 & 0.586509 \\
(3,11,4,10) & (8,3,7)   & -6 & 2.4791 & 3.12677 & 0.617705 & 0.59939 & 0.348088 & 0.434817 \\ \hline
(1,15,2,14) & (14,1,13) & -12 & 2.01893 & 3.83945 & 0.656246 & 0.645477 & 0.240967 & 0.45731 \\ 
(2,14,3,13) & (12,2,11) & -10 & 2.08323 & 3.72094 & 0.642184 & 0.629183 & 0.300885 & 0.427748 \\
(2,14,5,11) & (12,2,3)  & -2 & 2.10746 & 3.67816 & 0.643822 & 0.60599 & 0.243022 & 0.507166 \\
(3,13,5,11) & (10,3,4)  & -3 & 2.15875 & 3.59077 & 0.625837 & 0.595104 & 0.313746 & 0.465313 \\
(4,12,5,11) & (8,4,7)   & -6 & 2.19639 & 3.52923 & 0.606287 & 0.588973 & 0.369229 & 0.435511 \\
(5,11,7,9)  & (6,5,2)   & -1 & 2.25203 & 3.44204 & 0.582711 & 0.53978 & 0.392348 & 0.485161 \\ \hline
\end{tabular}
\end{small}
\vspace{0em} 
\begin{center}
Table 2: central charges and R-charges for $L^{p,q;r,s}$
\end{center}

\end{document}